\documentclass[aps,superscriptaddress,showpacs,pra,twocolumn,nofootinbib]{revtex4-2}
\usepackage[english]{babel}
\usepackage[dvipsnames]{xcolor}
\usepackage[normalem]{ulem}
\usepackage[colorlinks]{hyperref}
\hypersetup{colorlinks,linkcolor=red,citecolor=blue,urlcolor=blue,final}
\usepackage{mathtools,amsmath,amssymb,amsthm,bm,enumitem,array,float,mathrsfs,bbm,bbold}

\newcommand{\be}{\begin{equation}}
\newcommand{\ee}{\end{equation}}
\newcommand{\eqnref}[1]{Eq.~\eqref{#1}}
\newcommand{\figref}[1]{Fig.~\ref{#1}}
\newcommand{\secref}[1]{Sec.~\ref{#1}}
\newcommand{\appref}[1]{App.~\ref{#1}}
\newcommand{\tabref}[1]{Table~\ref{#1}}
\newcommand{\pare}[1]{\left( {#1} \right)}
\newcommand{\spare}[1]{\left[ {#1} \right]}

\newcommand{\pares}[1]{( {#1} )}
\newcommand{\spares}[1]{[ {#1} ]}

\newcommand{\mv}[1]{ \langle #1 \rangle }
\newcommand{\w}{\omega}
\newcommand{\wa}{\omega_{ \text{A}}}
\newcommand{\wi}{\omega_{ \text{I}}}
\newcommand{\wl}{\omega_{ \text{L}}}
\newcommand{\wxval}{2}
\newcommand{\wtvala}{50}
\newcommand{\wtvale}{1}
\newcommand{\etaG}{\eta}
\newcommand{\Gcm} {{\Gamma}_\mathrm{cm}}
\newcommand{\Grot}{{\Gamma}_\mathrm{rot}}
\newcommand{\inertia}{{I}}

\newcommand{\curlyH}{\mathcal{H}}

\newcommand{\rvec}{\mathbf{r}}
\newcommand{\rtvec}{\tilde{\mathbf{r}}}
\newcommand{\pvec}{\mathbf{p}}
\newcommand{\ptvec}{\tilde{\mathbf{p}}}
\newcommand{\Bvec}{\mathbf{B}}
\newcommand{\bvec}{\mathbf{b}}
\newcommand{\Fvec}{\mathbf{F}}
\newcommand{\Lvec}{\bm{L}}
\newcommand{\lvec}{\bm{\ell}}
\newcommand{\evec}{\mathbf{e}}
\newcommand{\mvec}{\mathbf{m}}
\newcommand{\dmvec}{\delta\mathbf{m}}
\newcommand{\ovec}{\bm{\omega}}
\newcommand{\oveceff}{\bm{\omega}_\text{eff} }
\newcommand{\Ovec}{\bm{\Omega}}
\newcommand{\muvec}{\bm{\mu}}

\newcommand{\dotrtvec}{\dot{\tilde{\mathbf{r}}}}

\newcommand{\dotptvec}{\dot{\tilde{\mathbf{p}}}}

\newcommand{\dotlvec}{\dot{\bm{\ell}}}
\newcommand{\dotOvec}{\dot{\mathbf{\Omega}}}
\newcommand{\dotmvec}{\dot{\mathbf{m}}}
\newcommand{\dotmp}{\dot{m}_\parallel}
\newcommand{\dotdmvec}{\delta\dot{\mathbf{m}}}

\newcommand{\dotevec}{\dot{\mathbf{e}}}

\begin{document}
\title{Stability of a Magnetically Levitated Nanomagnet in Vacuum: Effects of Gas and Magnetization Damping}

\author{Katja Kustura}
\affiliation{Institute for Quantum Optics and Quantum Information of the Austrian Academy of Sciences, A-6020 Innsbruck, Austria.}
\affiliation{Institute for Theoretical Physics, University of Innsbruck, A-6020 Innsbruck, Austria.}
\author{Vanessa Wachter}
\affiliation{Max Planck Institute for the Science of Light, Staudtstraße 2, 91058 Erlangen, Germany}
\affiliation{Department of Physics, University of Erlangen-Nürnberg, Staudtstraße 7, 91058 Erlangen, Germany}
\author{Adrián E. Rubio López}
\affiliation{Institute for Quantum Optics and Quantum Information of the Austrian Academy of Sciences, A-6020 Innsbruck, Austria.}
\affiliation{Institute for Theoretical Physics, University of Innsbruck, A-6020 Innsbruck, Austria.}
\author{Cosimo C. Rusconi}
\affiliation{Max-Planck-Institut f\"ur Quantenoptik, Hans-Kopfermann-Strasse 1, 85748 Garching, Germany.}
\affiliation{Munich Center for Quantum Science and Technology, Schellingstrasse 4, D-80799 M\"unchen, Germany.}

\date{\today}

\begin{abstract}
In the absence of dissipation a non-rotating magnetic nanoparticle can be stably levitated in a static magnetic field as a consequence of the spin origin of its magnetization. Here we study the effects of dissipation on the stability of the system, considering the interaction with the background gas and the intrinsic Gilbert damping of magnetization dynamics. At large applied magnetic fields we identify magnetization switching induced by Gilbert damping as the key limiting factor for stable levitation. At low applied magnetic fields and for small particle dimensions magnetization switching is prevented due to the strong coupling of rotation and magnetization dynamics, and the stability is mainly limited by the gas-induced dissipation. In the latter case, high vacuum should be sufficient to extend stable levitation over experimentally relevant timescales. Our results demonstrate the possibility to experimentally observe the phenomenon of quantum spin stabilized magnetic levitation. 
\end{abstract}

\maketitle

\section{Introduction}\label{sec:introduction}

The Einstein--de Haas~\cite{Einstein1915,Richardson1908} and Barnett effects~\cite{Barnett1915} are macroscopic manifestations of the internal angular momentum origin of magnetization: a change in the magnetization causes a change in the mechanical rotation and conversely. Because of the reduced moment of inertia of levitated nano- to microscale particles, these effects play a dominant role in the dynamics of such systems~\cite{Chudnovsky1994,Rusconi2016,Ganzhorn2016,Kusminskiy2016,Keshtgar2017,Stickler2021,Perdriat2021}. This offers the possibility to harness these effects for a variety of applications such as precise magnetometry~\cite{Kimball2016,Kumar2017,Band2018,Wang2019,Fadeev2021a,Fadeev2021b}, inertial sensing~\cite{PratCamps2017,Vinante2020}, coherent spin-mechanical control~\cite{Huillery2020,Gieseler2020}, and spin-mechanical cooling~\cite{Delord2020,Gonzalez-Ballestero2020} among others. Notable in this context is the possibility to stably levitate a ferromagnetic particle in a static magnetic field in vacuum~\cite{Rusconi2017a,Rusconi2017b}. Stable levitation is enabled by the internal angular momentum origin of the magnetization which, even in the absence of mechanical rotation, provides the required angular momentum to gyroscopically stabilize the system. Such a phenomenon, which we refer to as quantum spin stabilized levitation to distinguish it from the rotational stabilization of magnetic tops~\cite{Berry1996,Simon1997,Gov1999}, relies on the conservative interchange between internal and mechanical angular momentum. Omnipresent dissipation, however, exerts additional non-conservative torques on the system which might alter the delicate gyroscopic stability~\cite{Merkin2012,Simon1997}. It thus remains to be determined if stable levitation can be observed under realistic conditions, where dissipative effects cannot be neglected. 

In this article, we address this question. Specifically, we consider the dynamics of a levitated magnetic nanoparticle (nanomagnet hereafter) in a static magnetic field in the presence of dissipation originating both from the collisions with the background gas and from the intrinsic damping of magnetization dynamics (Gilbert damping)~\cite{Gilbert2004,Bertotti2009}, which are generally considered to be the dominant sources of dissipation for levitated nanomagnets~\cite{Xi2006,Keshtgar2017,Martinetz2018,Lyutyy2019,Band2018}. Confined dynamics can be observed only when the time over which the nanomagnet is levitated is longer than the period of center-of-mass oscillations in the magnetic trap.
When this is the case,  we define the system to be \emph{metastable}. We demonstrate that the system can be metastable in experimentally feasible conditions, with the levitation time and the mechanism behind the instability depending on the parameter regime of the system.  In particular, we show that at weak applied magnetic fields and for small particle dimensions (to be precisely defined below) levitation time can be significantly extended in high vacuum (i.e. pressures below $10^{-3}$ mbar). Our results evidence the potential of unambiguous experimental observation of quantum spin stabilized magnetic levitation. 

We emphasize that our analysis is particularly timely. Presently there is a growing interest in levitating and controlling magnetic systems in vacuum~\cite{Millen2020,GonzalezBallestero2021,Stickler2021}.
Current experimental efforts focus on levitation of charged paramagnetic ensembles in a Paul trap~\cite{Kuhlicke2014,Delord2018,Huillery2020}, diamagnetic particles in magneto-gravitational traps~\cite{Slezak2018,Zheng2020,Leng2021}, or ferromagnets above a superconductor~\cite{Wang2019,Timberlake2019,Gieseler2020}. Levitating ferromagnetic particles in a static magnetic trap offers a viable alternative, with the possibility of reaching larger mechanical trapping frequencies. 

The article is organized as follows. In \secref{sec:description} we introduce the model of the nanomagnet, and we define two relevant regimes for metastability, namely the atom phase and the Einstein--de Haas phase. In Sec. \ref{sec:atom} and \ref{sec:EdH} we analyze the dynamics in the atom phase and the Einstein--de Haas phase, respectively. We discuss our results in \secref{sec:discussion}. Conclusions and outlook are provided in \secref{sec:conclusion}. Our work is complemented by three appendices where we define the transformation between the body-fixed and laboratory reference frames (\appref{app:transformation}), analyze the effect of thermal fluctuations (\appref{app:noise}), and provide additional figures (\appref{app:additional}).

\section{Description of the system}\label{sec:description} 

\begin{figure}[t]
	\includegraphics[width=0.99\linewidth]{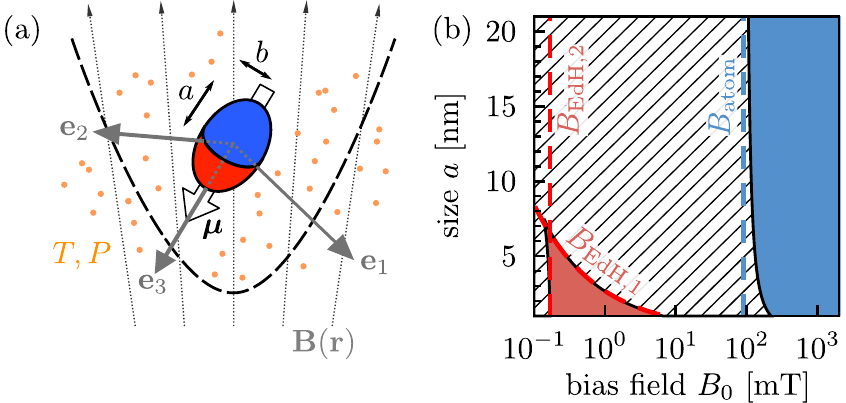}
	\caption{
		(a) Illustration of a spheroidal nanomagnet levitated in an external field $\Bvec(\rvec)$ and surrounded by a gas at the temperature $T$ and the pressure $P$. 
		(b) Linear stability diagram of a non-rotating nanomagnet in the absence of dissipation, assuming $a=2b$. Blue and red regions denote the stable atom and Einstein--de Haas phase, respectively; hatched area is the unstable region. Dashed lines show the critical values of the bias field which define the two phases. In particular, $B_\text{EdH,1} \equiv 5 \mu/\spares{4 \gamma_0^2 (a^2+b^2) M}$, $B_\text{EdH,2} \equiv 3\hspace{0.5mm}\spares{\mu B'^2/(4 \gamma_0 M) }^{1/3}$, and $B_\text{atom}=2k_aV/\mu$. Numerical values of physical parameters used to generate panel (b) are given in \tabref{tab:parameters}.  
		}
	\label{fig:model} 
\end{figure}

We consider a single domain nanomagnet levitated in a static\footnote{We denote a field \emph{static} if it does not have explicit time dependence, namely if $\partial \Bvec(\rvec) / \partial t = 0$.} magnetic field $\Bvec(\rvec)$ as shown schematically in \figref{fig:model}(a). We model the nanomagnet as a spheroidal rigid body of mass density $\rho_M$ and semi-axes lengths $a,b$ ($a>b$), having uniaxial magnetocrystalline anisotropy, with the anisotropy axis assumed to be along the major semi-axis $a$~\cite{Chikazumi2009}. Additionally, we assume that the magnetic response of the nanomagnet is approximated by a point dipole with magnetic moment $\muvec$ of constant magnitude $\mu \equiv \vert\muvec\vert$, as it is often justified for single domain particles~\cite{Chikazumi2009,Gatteschi2006}. Let us remark that such a simplified model has been considered before to study the classical dynamics of nanomagnets in a viscous medium~\cite{Newman1968,Tsebers1975,Scherer2000,Xi2006,Usadel2015,Usov2015,Lyutyy2018}, as well as to study the quantum dynamics of magnetic nanoparticles in vacuum~\cite{OKeeffe2011,OKeeffe2012,Rusconi2016,Band2018}. Since the model has been successful in describing the dynamics of single-domain nanomagnets, we adopt it here to investigate the stability in a magnetic trap. In particular, our study has three main differences as compared with previous work. (i) We consider a particle levitated in high vacuum, where the mean free path of the gas particles is larger than the nanomagnet dimensions (Knudsen regime~\cite{Cercignani1988}). This leads to gas damping which is generally different from the case of dense viscous medium mostly considered in the literature. (ii) We consider center-of-mass motion and its coupling to the rotational and magnetic degrees of freedom, while previous work mostly focuses on coupling between rotation and magnetization only (with the notable exception of~\cite{Usov2015}). (iii) We are primarily interested in the center-of-mass confinement of the particle, and not in its magnetic response. 

\begin{table}[t]
\caption{Physical parameters of the model and the values used throughout the article. We calculate the magnitude of the magnetic moment as $\mu = \rho_\mu V$, where $\rho_\mu = \rho_M \mu_B/(50\hspace{0.3mm}\text{amu})$, with $\mu_B$ the Bohr magneton and amu the atomic mass unit.}\label{tab:parameters}
\begin{ruledtabular}
\begin{tabular}{lll}
Parameter & Description & Value [units]\\
\hline
$\rho_M$ & mass density & $10^4$ [$\text{kg}\hspace{0.5mm} \text{m}^{-3}$] \\
$a,b\text{ }$ & semi-axes & see main text [$\text{m}$] \\
$\rho_\mu$ & magnetization & $2.2 \times 10^6$ [$\text{J}\hspace{0.5mm} \text{T}^{-1}\hspace{0.5mm} \text{m}^{-3}$]\\
$k_a$ & anisotropy constant & $10^5$ [$\text{J}\hspace{0.5mm} \text{m}^{-3}$]\\
$\gamma_0$ & gyromagnetic ratio & $1.76 \times 10^{11}$ [$\text{rad} \hspace{0.5mm}\text{s}^{-1}\hspace{0.5mm}\text{T}^{-1}$]\\
$B_0$ & field bias & see main text [$\text{T}$]\\
$B'$ & field gradient & $10^4$ [$\text{T}\hspace{0.5mm}\text{m}^{-1}$]\\
$B''$ & field curvature & $10^6$ [$\text{T}\hspace{0.5mm}\text{m}^{-2}$]\\
$\etaG$ & Gilbert damping & $10^{-2}$ [n. u.]\\
$T$ & temperature & $10^{-1}$ [$\text{K}$] \\
$P$ & pressure & $10^{-2}$ [$\text{mbar}$]\\
$\mathcal{M}$ & molar mass & $29$ [$\text{g}\hspace{0.5mm} \text{mol}^{-1}$]\\
$\alpha_c$ & reflection coefficient & 1 [n. u.]\\
\end{tabular}
\end{ruledtabular}
\end{table}

Within this model the relevant degrees of freedom of the system are the center-of-mass position $\rvec$, the linear momentum $\pvec$, the mechanical angular momentum $\Lvec$, the orientation of the nanomagnet in space $\Ovec$, and the magnetic moment $\muvec$. The orientation of the nanomagnet is specified  by the body-fixed reference frame $O\evec_1\evec_2\evec_3$, which is obtained from the laboratory frame $O\evec_x\evec_y\evec_z$ according to $(\evec_1,\evec_2,\evec_3)^T=R(\Ovec)(\evec_x,\evec_y,\evec_z)^T$, where $\Ovec = (\alpha,\beta,\gamma)^T$ are the Euler angles and $R(\Ovec)$ is the rotational matrix. We provide the expression for $R(\Ovec)$ in~\appref{app:transformation}. The body-fixed reference frame is chosen such that $\evec_3$ coincides with the anisotropy axis. The magnetic moment $\muvec$ is related to the internal angular momentum $\Fvec$ according to the gyromagnetic relation $\muvec=\gamma_0 \Fvec $, where $\gamma_0$ is the gyromagnetic ratio of the material\footnote{The total internal angular momentum $\Fvec$ is a sum of the individual atomic angular momenta (spin and orbital), which contribute to the atomic magnetic moment. For a single domain magnetic particle, it is customary to assume that $\Fvec$ can be described as a vector of constant magnitude, $\vert \Fvec\vert = \mu/\gamma_0$ (macrospin approximation)~\cite{Gatteschi2006}.}.

\subsection{Equations of Motion}\label{sec:EoM}

We describe the dynamics of the nanomagnet in the magnetic trap with a set of stochastic differential equations which model both the deterministic dissipative evolution of the system and the random fluctuations due to the environment. In the following it is convenient to define dimensionless variables: the center-of-mass variables $\rtvec\equiv\rvec/a$, $\ptvec\equiv \gamma_0 a\pvec/\mu$, the mechanical angular momentum $\lvec \equiv \gamma_0 \Lvec/\mu$, the magnetic moment $\mvec \equiv \muvec/\mu$, and the magnetic field $\bvec(\rtvec)\equiv\Bvec(a\rtvec)/B_0$, where $B_0$ denotes the minimum of the field intensity in a magnetic trap, which we hereafter refer to as the bias field. Note that we choose to normalize the position $\rvec$, the magnetic moment $\muvec$ and the magnetic field $\Bvec(\rvec)$ with respect to the particle size $a$, the magnetic moment magnitude $\mu$, and the bias field $B_0$, respectively. The scaling factor for angular momentum, $\mu/\gamma_0$, and linear momentum, $\mu/(a\gamma_0)$, follow as a consequence of the gyromagnetic relation. 

The dynamics of the nanomagnet in the laboratory frame are given by the equations of motion 
\begin{align}
	\dotrtvec &= \wi \ptvec,\label{eom-r} \\
	\dotevec_3  &= \ovec \times \evec_3\vphantom{\frac{1}{2}},\label{eom-e}\\
	\dotptvec &= \wl \nabla_{\rtvec} [\mvec \cdot \bvec(\rtvec)]- \Gcm \ptvec + \bm{\phi}_p(t),\label{eom-p}\\
	\dotlvec  &= \wl \mvec \times \bvec(\rtvec) -\dotmvec - \Grot\lvec + \bm{\xi}_l(t), \label{eom-l}\\
	\dotmvec  &= \frac{\mvec}{1+\etaG^2}\!  \times\! [\oveceff - \etaG\mvec\times\pare{\ovec+\oveceff+ \etaG \ovec \times \mvec  }\nonumber\\
	& +\bm{\zeta}_b(t)].\label{eom-m}
\end{align}
Here we define the relevant system frequencies: $\wi \equiv \mu / (\gamma_0 M a^2)$ is the Einstein--de Haas frequency, with $M$ the mass of the nanomagnet, $\wl \equiv \gamma_0 B_0$ is the Larmor frequency, $\wa \equiv k_a V \gamma_0/\mu$ is the anisotropy frequency, with  $V$ the volume of the nanomagnet and $k_a$ the material dependent anisotropy constant~\cite{Gatteschi2006}, $\ovec \equiv \inertia^{-1}\Lvec$ is the angular velocity, with $\inertia$ the tensor of inertia, and $\oveceff\equiv 2 \wa (\mvec \cdot \evec_3) \evec_3 + \wl \bvec(\rtvec)$. Dissipation is parametrized by the dimensionless Gilbert damping parameter $\etaG$~\cite{Gilbert2004,Miltat2002}, and the center-of-mass and rotational friction tensors $\Gcm$ and $\Grot$, respectively~\cite{Martinetz2018}. The effect of stochastic thermal fluctuations is represented by the random variables $\bm{\phi}_p(t)$ and $\bm{\xi}_{l}(t)$ which describe, respectively, the fluctuating force and torque exerted by the surrounding gas, and by $\bm{\zeta}_{b}(t)$ which describes the random magnetic field accounting for thermal fluctuations in magnetization dynamics~\cite{Brown1963}. We assume Gaussian white noise, namely, for $\bm{X}(t)\equiv(\bm{\phi}_p(t),\bm{\xi}_l(t),\bm{\zeta}_b(t))^T$ we have $\mv{X_i(t)} = 0$ and $\mv{X_i(t)X_j(t')} \sim \delta_{ij}\delta(t-t')$.  

Equations~(\ref{eom-r}-\ref{eom-l}) describe the center-of-mass and rotational dynamics of a rigid body in the presence of dissipation and noise induced by the background gas~\cite{Martinetz2018}. The expressions for $\Gcm$ and $\Grot$ depend on the particle shape -- here we take the expressions derived in~\cite{Martinetz2018} for a cylindrical particle\footnote{The expressions for $\Gcm$ and $\Grot$ for a cylindrical particle capture the order of magnitude of the dissipation rates for a spheroidal particle~\cite{Martinetz2020,Schaefer2021}.} --, and on the ratio of the surface and the bulk temperature of the particle, which we assume to be equal to the gas temperature $T$. Furthermore, they account for two different scattering processes, namely the specular and the diffusive reflection of the gas from the particle, which is described by a phenomenological interpolation coefficient $\alpha_c$. The order of magnitude of the different components of $\Gcm$ and $\Grot$ is generally well approximated by the dissipation rate $\Gamma \equiv ( 2 P a b / M) \spares{2 \pi \mathcal{M} /(N_A k_B T)}^{1/2}$, where $P$ and $\mathcal{M}$ are, respectively, the gas pressure and molar mass, $k_B$ is the Boltzmann constant and $N_A$ is the Avogadro number. The magnetization dynamics given by \eqnref{eom-m} is the Landau-Lifshitz-Gilbert equation in the laboratory frame~\cite{Taylor2005,Keshtgar2017}, with the effective magnetic field $\oveceff/\gamma_0$. We remark that Eqs.~(\ref{eom-r}-\ref{eom-m}) describe the classical dynamics of a levitated nanomagnet where the effect of the quantum spin origin of magnetization, namely the gyromagnetic relation, is taken into account phenomenologically by \eqnref{eom-m}. This is equivalent to the equations of motion obtained from a quantum Hamiltonian in the mean-field approximation~\cite{Rusconi2017b}. 

Let us discuss the effect of thermal fluctuations on the dynamics of the nanomagnet at subkelvin temperatures and in high vacuum. These conditions are common in recent experiments with levitated particles~\cite{Delic2020,Magrini2021,Tebbenjohanns2021}. The thermal fluctuations of magnetization dynamics, captured by the last term in \eqnref{eom-m}, lead to thermally activated transition of the magnetic moment between the two stable orientations along the anisotropy axis~\cite{Brown1963,Shliomis1974}. Such process can be quantified by the N\'{e}el relaxation time, which is given by $\tau_\text{N} \approx (\pi/\wa)\sqrt{k_B T /(k_a V)} e^{k_a V/(k_B T )}$. Thermal activation can be neglected when $\tau_\text{N}$ is larger than other timescales of magnetization dynamics, namely the precession timescale given by $\tau_\text{L} \equiv 1/\vert\oveceff\vert$, and the Gilbert damping timescale given by $\tau_\text{G} \equiv 1/(\etaG \vert\oveceff\vert)$. Considering for simplicity $\vert\oveceff\vert \sim 2 \wa$, for a particle size $a=2b=1$ nm and temperature $T=1$ K, and the values of the remaining parameters as in \tabref{tab:parameters}, the ratio of the timescales is of the order $\tau_\text{N}/\tau_\text{L} \sim 10^3$, and it is significantly increased for larger particle sizes and at smaller temperatures. We remark that, for the values considered in this article, $\tau_\text{N}$ is much larger than the longest dynamical timescale in Eqs.~(\ref{eom-r}-\ref{eom-m}) which is associated with the motion along $\evec_x$. Thermal activation of the magnetic moment can therefore be safely neglected. The stochastic effects ascribed to the background gas, captured by the last terms in Eqs.~(\ref{eom-p}-\ref{eom-l}), are expected to be important at high temperatures (namely, a regime where $M k_B T \gamma_0^2 a^2/\mu^2\gtrsim 1$~\cite{Martinetz2018}). At subkelvin temperatures and in high vacuum these fluctuations are weak and, consequently, they do not destroy the deterministic effects captured by the remaining terms in Eqs. (\ref{eom-r}-\ref{eom-m})~\cite{Lyutyy2019}. Indeed, for the values of parameters given in \tabref{tab:parameters} and for $a=2b$, $M k_B T \gamma_0^2 a^2/\mu^2 \approx 0.8 \hspace{0.5mm}T / (a\text{[nm]})$. For subkelvin temperatures and particle sizes $a > 1 $ nm, thermal fluctuations due to the background gas can therefore be safely neglected. 

In the following we thus neglect stochastic effects by setting $\bm{\phi}_p=\bm{\xi}_l=\bm{\zeta}_b=0$, and we consider only the deterministic part of Eqs.~(\ref{eom-r}-\ref{eom-m}) as an appropriate model for the dynamics~\cite{Lyutyy2019,Keshtgar2017,Brown1963}. In \appref{app:noise} we carry out the analysis of the dynamics including the effects of gas fluctuations in equations~(\ref{eom-r}-\ref{eom-m}), and we show that the results presented in the main text remain qualitatively valid even in the presence of thermal noise. For the magnetic field $\Bvec(\rvec)$ we hereafter consider a Ioffe-Pritchard magnetic trap, given by 
\be\label{ioffe-pritchard}
\begin{split}
\Bvec(\rvec) =& \,
 \evec_x \spare{B_0 + \frac{B''}{2}\pare{x^2-\frac{y^2+z^2}{2}}}\\
&-\evec_y \pare{B'y + \frac{B''}{2}xy }+\evec_z \pare{B'z - \frac{B''}{2}xz },
\end{split}
\ee
where $B_0,B'$ and $B''$ are, respectively, the field bias, gradient and curvature~\cite{Reichel2011}. We remark that this is not a fundamental choice, and different magnetic traps, provided they have a non-zero bias field, should result in similar qualitative behavior.

\subsection{Initial conditions}\label{sec:initial} 

The initial conditions for the dynamics in Eqs.~(\ref{eom-r}-\ref{eom-m}), namely at time $t=0$, depend on the initial state of the system, which is determined by the preparation of the nanomagnet in the magnetic trap.  In our analysis, we consider the nanomagnet to be prepared in the thermal state of an auxiliary loading potential at the temperature $T$. Subsequently, we assume to switch off the loading potential at $t=0$, while at the same time switching on the Ioffe-Pritchard magnetic trap. The choice of the auxiliary potential is determined by two features: (i) it allows us to simply parametrize the initial conditions by a single parameter, namely the temperature $T$, and (ii) it is an adequate approximation of general trapping schemes used to trap magnetic particles.

Regarding point (i), we assume that the particle is levitated in a harmonic trap, in the presence of an external magnetic field applied along $\evec_x$. This loading scheme provides, on the one hand, trapping of the center-of-mass degrees of freedom, with trapping frequencies denoted by $\w_i$ ($i=x,y,z$). On the other hand, the magnetic moment in this case is polarized along $\evec_x$. The Hamiltonian of the system in such a configuration reads $\curlyH_\text{aux}= \pvec^2/(2M) + \sum_{i=x,y,z} M\w_i^2 r_i^2/2 + \Lvec \inertia^{-1} \Lvec/2 - k_a V e_{3,x}^2 -\mu_x B_\text{aux}$, where $B_\text{aux}$ denotes the magnitude of the external magnetic field, which we for simplicity set to $B_\text{aux}=B_0$ in all our simulations. At $t=0$ the particle is released in the magnetic trap given by \eqnref{ioffe-pritchard}. For the degrees of freedom $\bm{x}\equiv(\rtvec,\ptvec,\lvec,m_x)^T$, we take as the initial displacement from the equilibrium the corresponding standard deviation in a thermal state of $\curlyH_\text{aux}$. More precisely, $x_i(0) = x_{i,e}+(\mv{x_i^2}-\mv{x_i}^2)^{1/2}$, where $x_{i,e}$ denotes the equilibrium value, and $\mv{x_i^k}\equiv Z^{-1}\int \text{d}  \bm{x} \hspace{1mm}x_i^k \exp[-\curlyH_\text{aux}/(k_B T)] $, with $k=1,2$ and the partition function $Z$. For the Euler angles $\Ovec$ we use $\Omega_1 (0) \equiv \cos^{-1} \spares{-\sqrt{\mv{\cos^2\Omega_1 }}}$ and $\Omega_{i} (0)\equiv \cos^{-1} \spares{\sqrt{\mv{\cos^2\Omega_i}}}$ ($i=2,3$). The initial conditions for $\evec_3$ follow from $\Ovec$ using the transformation given in \appref{app:transformation}. 

Regarding point (ii), the initial conditions obtained in this way describe a trapped particle prepared in a thermal equilibrium in the presence of an external loading potential where the center of mass is decoupled from the magnetization and the rotational dynamics. It is outside the scope of this article to study in detail a particular loading scheme. However, we point out that an auxiliary potential given by $\curlyH_\text{aux}$ can be obtained, for example, by trapping the nanomagnet using a Paul trap as demonstrated in recent experiments~\cite{Millen2015,Alda2016,Conangla2018,Delord2018,Ostermayr2018,Partner2018,Bykov2019,Conangla2020,Delord2020,Huillery2020,Dania2021}. In particular, trapping of a ferromagnetic particle has been demonstrated in a Paul trap at $P=10^{-2}~\text{mbar}$, with center-of-mass trapping frequency of up to $1~\text{MHz}$, and alignment of the particle along the direction of an applied field~\cite{Huillery2020}. We note that particles are shown to remain trapped even when the magnetic field is varied over many orders of magnitudes or switched off. We remark further that alignment of elongated particles can be achieved using a quadrupole Paul trap even in the absence of magnetic field ~\cite{Martinetz2020,Martinetz2021}.

\subsection{Linear stability}\label{sec:phases} 

In the absence of thermal fluctuations, an equilibrium solution of Eqs.~(\ref{eom-r}-\ref{eom-m}) is given by $\rtvec_e=\ptvec_e=\lvec_e=0$ and $\evec_{3,e}=\mvec_e=-\evec_x$. This corresponds to the configuration in which the nanomagnet is fixed at the trap center, with the magnetic moment along the anisotropy axis and anti-aligned to the bias field $B_0$. Linear stability analysis of Eqs.~(\ref{eom-r}-\ref{eom-m}) shows that the system is unstable, as expected for a gyroscopic system in the presence of dissipation~\cite{Merkin2012}. However, when the nanomagnet is metastable, it is still possible for it to levitate for an extended time before being eventually lost from the trap, as in the case of a classical magnetic top~\cite{Berry1996,Simon1997,Gov1999}. As we show in the following sections, the dynamics of the system, and thus its metastability, strongly depend on the applied bias field $B_0$. We identify two relevant regimes: (i) strong-field regime, defined by bias field values $B_0>B_\text{atom}$, and (ii) weak-field regime, defined by $B_0<B_\text{atom}$,  where $B_\text{atom} \equiv 2 k_a V/\mu$. This difference is reminiscent of the two different stable regions which arise as a function of $B_0$ in the linear stability diagram in the \emph{absence} of dissipation [see~\figref{fig:model}(b)]~\cite{Rusconi2017a,Rusconi2017b}. In \secref{sec:atom} and \secref{sec:EdH} we investigate the possibility of metastable levitation by solving numerically Eqs. (\ref{eom-r}-\ref{eom-m}) in the strong-field  and weak-field regime, respectively. 

\section{Dynamics in the strong-field regime: atom phase}\label{sec:atom}

The strong-field regime, according to the definition given in \secref{sec:phases}, corresponds to the blue region in the linear stability diagram in the absence of dissipation, shown in~\figref{fig:model}(b). This region is named \emph{atom phase} in~\cite{Rusconi2017a,Rusconi2017b}, and we hereafter refer to the strong-field regime as the atom phase. This parameter regime corresponds to the condition $\wl \gg \wa,\wi$. In this regime, the coupling of the magnetic moment $\muvec$ and the anisotropy axis $\evec_3$ is negligible, and, to first approximation, the nanomagnet undergoes a free Larmor precession about the local magnetic field. In the absence of dissipation, this stabilizes the system in full analogy to magnetic trapping of neutral atoms~\cite{Sukumar1997,Brink2006}. 

\begin{figure}[t]
	\includegraphics[width=0.99\linewidth]{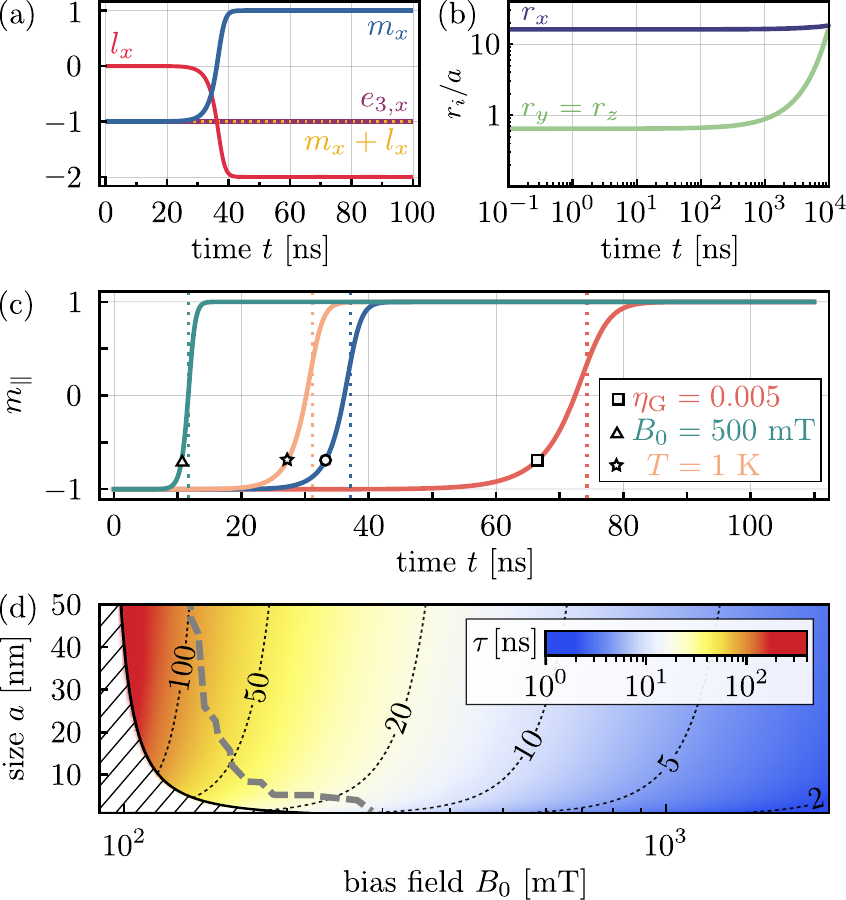}
	\caption{Dynamics in the atom phase. (a) Dynamics of the magnetic moment component $m_x$, the mechanical angular momentum component $l_x$, and the anisotropy axis component $e_{3,x}$ for nanomagnet dimensions $a=2b=20$ nm and the bias field $B_0=200$~mT. For the initial conditions we consider trapping frequencies $\w_x = 2\pi \times \wxval$ kHz and $\w_y=\w_z = 2\pi \times \wtvala$ kHz. Unless otherwise stated, for the remaining parameters the numerical values are given in \tabref{tab:parameters}. 
	(b) Center-of-mass dynamics for the same case considered in (a). 
	(c) Dynamics of the magnetic moment component $m_\parallel$. Line denoted by circle corresponds to the case considered in (a). Each remaining line differs by a single parameter, as denoted by the legend. Dotted vertical lines show \eqnref{tau}. 
	(d) Switching time given by \eqnref{tau} as a function of the bias field $B_0$ and the major semi-axis $a$. In the region left of the thick dashed line the deviation from the exact value is more than $5 \%$. Hatched area is the unstable region in the linear stability diagram in \figref{fig:model}.(b). }
	\label{fig:atom} 
\end{figure}

In \figref{fig:atom}(a-c) we show the numerical solution of Eqs.~(\ref{eom-r}-\ref{eom-m}) for nanomagnet dimensions $a=2b=20$ nm and the bias field $B_0 = 200$ mT. As evidenced by \figref{fig:atom}(a), the magnetization $m_x$ of the particle changes direction. During this change, the mechanical angular momentum $l_x$ changes accordingly in the manifestation of the Einstein--de Haas effect, such that the total angular momentum $\mvec+\lvec$ is conserved\footnote{We always find the transfer of angular momentum to the center of mass angular momentum $\rvec\times \pvec$ to be negligible.}. The dynamics observed in \figref{fig:atom}(a) is indicative of Gilbert-damping-induced magnetization switching, a well-known phenomenon in which the projection of the magnetic moment along the effective magnetic field $\oveceff/\gamma_0$ changes sign~\cite{Bertotti2009}. This is expected to happen when the applied bias field $B_0$ is larger than the effective magnetic field associated with the anisotropy, given by $\sim\wa/\gamma_0$. Magnetization switching displaces the system from its equilibrium position on a timescale which is much shorter than the period of center-of-mass oscillations, estimated from~\cite{Rusconi2017b} to be $\tau_\text{cm} \sim 1$ $\mu$s. The nanomagnet thus shows no signature of confinement [see~\figref{fig:atom}(b)].  

The timescale of levitation in the atom phase is given by the timescale of magnetization switching, which we estimate as follows. As evidenced by \figref{fig:atom}(a-b), the dynamics of the center of mass and the anisotropy axis are approximately constant during switching, such that $\oveceff \approx \oveceff(t=0)$. Under this approximation and assuming $\etaG \ll 1$, the magnetic moment projection $m_\parallel \equiv \oveceff \cdot\mvec /\vert \oveceff \vert$ evolves as
\be\label{magn-sw}
\begin{split}
\dotmp \approx	\etaG	\spares{\wl+{2\wa}m_\parallel} \pares{ 1 -m_\parallel^2 }.   
\end{split}
\ee
According to \eqnref{magn-sw} the component $m_\parallel$ exhibits switching if $m_\parallel(t=0)\gtrsim-1$ and $\wl/2\wa>1$~\cite{Bertotti2009}, both of which are fulfilled in the atom phase. Integrating \eqnref{magn-sw} we obtain the switching time $\tau$ [defined as $m_\parallel(\tau)\equiv0$], which can be well approximated by 
\be\label{tau}
\begin{split}
	\tau \approx
	 \frac{ \ln \pare{1+ \vert m_\parallel(t=0)\vert } }{2\etaG\pare{\wl+2\wa}}
	-\frac{ \ln \pare{1- \vert m_\parallel(t=0)\vert } }{2\etaG\pare{\wl-2\wa}}.
\end{split}
\ee
The estimation \eqnref{tau} is in excellent agreement with the numerical results for different parameter values [see \figref{fig:atom}(c)]. 

\begin{figure*}[t]
	\includegraphics[width=0.99\linewidth]{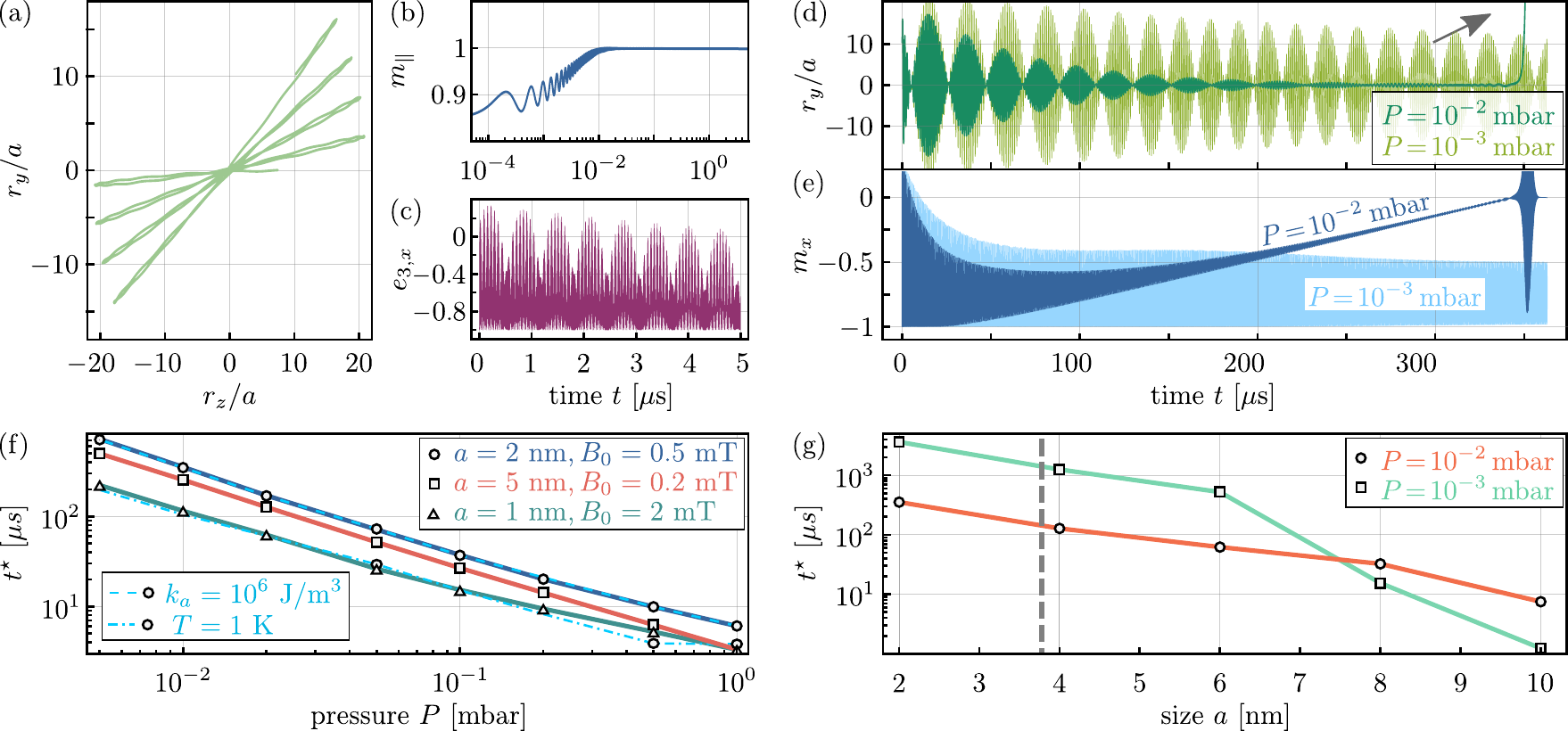}
	\caption{Dynamics in the Einstein--de Haas phase. (a) Motion of the system in the $\evec_y$-$\evec_z$ plane until time $t = 5$~$\mu$s for nanomagnet dimensions $a=2b=2$ nm and the bias field $B_0=0.5$~mT. For the initial conditions we consider trapping frequencies $\w_x = 2\pi \times \wxval$ kHz and $\w_y=\w_z = 2\pi \times \wtvale$ MHz. For the remaining parameters the numerical values are given in \tabref{tab:parameters}. (b) Dynamics of the projection $m_\parallel$ and (c) dynamics of the anisotropy axis component $e_{3,x}$, for the same case considered in (a). (d) Dynamics of the center-of-mass component $r_y$ and (e) dynamics of the magnetic moment component $m_x$ on a longer timescale, for the same values of parameters as in (a). (f) Escape time $t^\star$ as a function of gas pressure $P$, for different configurations in the Einstein--de Haas phase. Circles correspond to the case considered in (a). Each remaining case differs by parameters indicated by the legend. (g) Escape time $t^\star$ as a function of the major semi-axis $a$, with the values of the remaining parameters as in (a). Dashed vertical line denotes the upper limit of the Einstein--de Haas phase, given by the critical field $B_\text{EdH,1}$ [see \figref{fig:model}(b)]. }
	\label{fig:EdH} 
\end{figure*}

Magnetization switching characterizes the dynamics of the system in the entire atom phase. In particular, in \figref{fig:atom}(d) we analyze the validity of \eqnref{tau} for different values of the bias field $B_0$ and the major semi-axis $a$, assuming $b=a/2$. The thick dashed line shows the region where \eqnref{tau} differs from the exact switching time, as estimated from the full dynamics of the system,  by $5 \%$; left of this line the deviation becomes increasingly more significant, with \eqnref{tau} predicting up to $20 \%$ larger values close to the stability border (namely, for bias field close to $B_\text{atom}=90$ mT). We believe that the significant deviation close to the border of the atom phase is due to the non-negligible coupling to the anisotropy axis, which results in additional mechanisms not captured by the simple model \eqnref{magn-sw}. In fact, it is known that coupling between magnetization and mechanical degrees of freedom might have an impact on the switching dynamics~\cite{Kovalev2005}. As demonstrated by \figref{fig:atom}(d), the switching time is always shorter than the center-of-mass oscillation period $\tau_\text{cm}$, and thus no metastability can be observed in the atom phase.

Let us note that the conclusions we draw in~\figref{fig:atom} remain valid if one varies the anisotropy constant $k_a$, Gilbert damping parameter $\etaG$, and the temperature $T$, as we show in~\appref{app:additional}. Finally, we note that the dissipation due to the background gas has negligible effects. In particular, for the values assumed in~\figref{fig:atom}(a-b) the timescale of the gas-induced dissipation is given by $1/\Gamma=440$ $\mu$s.

\section{Dynamics in the weak-field regime: Einstein--de Haas phase}\label{sec:EdH}

We now focus on the regime of weak bias field, corresponding to the condition
$\wl \ll \wa$. In this regime magnetization switching does not occur, and the dynamics critically depend on the particle size. In the following we focus on the regime of small particle dimensions, i.e. $\wl \ll \wi$, which, as we will show, is beneficial for metastability. In the absence of dissipation, this regime corresponds to the \emph{Einstein--de Haas phase} [red region in~\figref{fig:model}(b)]~\cite{Rusconi2017a,Rusconi2017b}. The hierarchy of energy scales in the Einstein--de Haas phase (namely, $\wl \ll \wa,\wi$) manifests in two ways: (i) the anisotropy is strong enough to effectively ``lock'' the direction of the magnetic moment $\muvec$ along the anisotropy axis $\evec_3$ ($\wa\gg \wl$), and (ii) according to the Einstein--de Haas effect, the frequency at which the nanomagnet would rotate if $\muvec$ switched direction is significantly increased at small dimensions ($\wi\gg \wl$), such that switching can be prevented due to energy conservation~\cite{Chudnovsky1994}. In the absence of dissipation, the combination of these two effects stabilizes the system. 

In \figref{fig:EdH}(a-c) we show the numerical solution of Eqs.~(\ref{eom-r}-\ref{eom-m}) for nanomagnet dimensions $a=2b=2$ nm and the bias field $B_0 = 0.5$ mT. The nanomagnet is metastable, as evidenced by the confined center-of-mass motion shown in \figref{fig:EdH}(a). In \figref{fig:EdH}(b-c) we show the dynamics of the magnetic moment component $m_\parallel$ and the anisotropy axis component $\evec_{3,x}$, respectively, which indicates that no magnetization switching occurs in this regime. We remark that the absence of switching cannot be simply explained on the basis of Eqs.~(\ref{magn-sw}-\ref{tau}). In fact, the simple model of magnetization switching, given by \eqnref{magn-sw}, assumes that the dynamics of the rotation and the center-of-mass motion happen on a much longer timescale than the timescale of magnetization dynamics. However, in this case rotation and magnetization dynamics occur on a comparable timescale, as evidenced by \figref{fig:EdH}(b-c). The weak-field condition alone ($\wl\ll\wa$) is thus not sufficient to correctly explain the absence of switching, and the role of particle size ($\wl\ll\wi$) needs to be considered.

Let us analyze the role of Gilbert damping in this case. Since in the Einstein--de Haas phase $m_\parallel \sim 1$, we define $\mvec \equiv \evec_3 + \dmvec$, where $\dmvec$ represents the deviation of $\mvec$ from the anisotropy axis $\evec_3$, and we assume $\vert\dmvec\vert \ll \vert \evec_3 \vert$ [see \figref{fig:EdH}(b)]. This allows us to simplify \eqnref{eom-m} as
\be\label{correction}
\begin{split}
	\dotdmvec \approx \oveceff\times\dmvec - \etaG\spare{ 2\wa + \w_3 \evec_3\cdot (\mvec+\lvec) } \dmvec,
\end{split}
\ee
where $\omega_3 \equiv \mu/(\gamma_0 I_3)$, with $I_3$ the principal moment of inertia along $\evec_3$. As evidenced by \eqnref{correction}, the \emph{only} effect of Gilbert damping is to align $\mvec$ and $\evec_3$ on a timescale given by $\tau' \equiv 1/[\etaG (2\wa+\w_3)]$, irrespective of the dynamics of $\evec_3$. For the values of parameters considered in \figref{fig:EdH}(a-c), $\tau'=5$ ns, and it is much shorter than the timescale of center-of-mass dynamics, given by $\tau_\text{cm}\sim 1$ $\mu$s. For all practical purposes, the magnetization in the Einstein--de Haas phase can be considered frozen along the anisotropy axis. The nanomagnet in the presence of Gilbert damping is therefore equivalent to a \emph{hard magnet} (i. e. $k_a \to \infty$)~\cite{Rusconi2017b}. 

The main mechanism behind the instability in the Einstein--de Haas phase is thus gas-induced dissipation. In \figref{fig:EdH}(d-e) we plot the dynamics of the center-of-mass component $r_y$ and the magnetic moment component $m_x$ on a longer timescale, for two different values of the pressure $P$. The effect of gas-induced dissipation is to dampen the center-of-mass motion to the equilibrium position, while the magnetic moment moves away from the equilibrium. Both processes happen on a timescale given by the dissipation rate $\Gamma$. When $e_x = m_x \approx 0$, the system becomes unstable and ultimately leaves the trap [see arrow in \figref{fig:EdH}(d)]. We define the \emph{escape time} $t^\star$ as the time at which the particle position is $y(t^\star)\equiv 5 y(0)$, and we show it in \figref{fig:EdH}(f) as a function of pressure $P$ for different configurations in the Einstein--de Haas phase, and for $b=a/2$. \figref{fig:EdH}(f) confirms that the dissipation affects the system on a timescale which scales as $\sim 1/P$. The metastability of the nanomagnet in the Einstein--de Haas phase is therefore limited solely by the gas-induced dissipation, which can be significantly reduced in high vacuum. Finally, in \figref{fig:EdH}(g) we analyze the effect of particle size on metastability. Specifically, we show the escape time $t^\star$ as a function of the major semi-axis $a$ at the bias field $B_0 = 0.5$ mT, for $b=a/2$. The escape time is significantly reduced at increased particle sizes. This confirms the advantage of the Einstein--de Haas phase to observe metastability, even in the presence of dissipation. 

\section{Discussion}\label{sec:discussion}

In deriving the results discussed in the preceding sections, we assumed (i) a single-magnetic-domain nanoparticle with uniaxial anisotropy and constant magnetization, with the values of the physical parameters summarized in \tabref{tab:parameters}, (ii) deterministic dynamics, i. e. the absence of thermal fluctuations, (iii) that gravity can be neglected, and (iv) a non-rotating nanomagnet. Let us justify the validity of these assumptions. 

We first discuss the values of the parameters given in \tabref{tab:parameters}, which are used in our analysis. The material parameters, such as $\rho_M$, $\rho_\mu$, $k_a$ and $\etaG$, are consistent with, for example, cobalt~\cite{Cullity2008,Walowski2008,Barati2014,Papusoi2018}. We remark that the uniaxial anisotropy considered in our model represents a good description even for materials which do not have an intrinsic magnetocrystalline uniaxial anisotropy, provided that they have a dominant contribution from the uniaxial \emph{shape} anisotropy. This is the case, for example, for ferromagnetic particles with a prolate shape~\cite{Cullity2008}. We point out that the values used here do not correspond to a specific material, but instead they describe a general order of magnitude corresponding to common magnetic materials. Indeed, our results are general and can be particularized to specific materials by replacing the above generic values with exact numbers. As we show in \appref{app:additional}, the results and conclusions presented here remain unchanged even when different values of the parameters are considered. The values used for the field gradient $B'$ and the curvature $B''$ have been obtained in magnetic microtraps~\cite{Reichel2001,Reichel2002,Barb2005,Fortagh2007,Reichel2011}. The values of the gas pressure $P$ and the temperature $T$ are experimentally feasible, with numerous recent experiments reaching pressure values as low as $P = 10^{-6}$ mbar~\cite{Bykov2019,Meyer2019,Windey2019,Delic2020,Dania2021,Sommer2021}. All the values assumed in our analysis are therefore consistent with currently available technologies in levitated optomechanics. 

Thermal fluctuations can be neglected at cryogenic conditions (as we argue in \secref{sec:EoM}), as their effect is weak enough not to destroy the deterministic effects captured by Eqs. (\ref{eom-r}-\ref{eom-m}). In particular, thermal activation of the magnetization, as quantified by the N\'{e}el relaxation time, can be safely neglected due to the large value of the uniaxial anisotropy even for the smallest particles considered. As for the mechanical thermal fluctuations, we confirm that they do not modify the deterministic dynamics in \appref{app:noise}, where we simulate the associated stochastic dynamics.  

Gravity, assumed to be along $\evec_x$, can be safely neglected, since the gravity-induced displacement of the trap center from the origin is much smaller than the length scale over which the Ioffe-Pritchard field significantly changes~\cite{Rusconi2017b}. Specifically, the gravitational potential $M g x$ shifts the trap center from the origin $\rvec = 0$ along $\evec_x$ by an amount $r_g \equiv M g/(\mu B'')$, where $g$ is the gravitational acceleration. On the other hand, the characteristic length scales of the Ioffe-Pritchard field are given by $\Delta r_0 \equiv \sqrt{B_0/B''}$ for the variation along $\evec_x$, and $\Delta r' \equiv B'/B''$ for the variation off-axis. Whenever $r_g\ll \Delta r_0,\Delta r'$, gravity has a negligible role in the metastable dynamics of the system. In the parameter regime considered in this article, this is always the case. We note that the condition to neglect gravity is the same as for a magnetically trapped atom, since both $M$ and $\mu$ scale with the volume.  

Finally, we remark that the analysis presented here is carried out for the case of a non-rotating nanomagnet\footnote{Rotational cooling might be needed to unambiguously identify the internal spin as the source of stabilization. Subkelvin cooling of a nanorotor has been recently achieved~\cite{VanderLaan2020,VanderLaan2021}, and cooling to $\mu$K temperatures should be possible~\cite{Schaefer2021}.}. The same qualitative behavior is obtained even in the presence of mechanical rotation (namely, considering a more general equilibrium configuration with $\lvec_e \neq 0$). The analysis of dynamics in the presence of rotation is provided in~\appref{app:additional}. In particular, the dynamics in the Einstein--de Haas phase remains largely unaffected, provided that the total angular momentum of the system is not zero. In the atom phase, mechanical rotation leads to differences in the switching time $\tau$, as generally expected in the presence of magneto-mechanical coupling~\cite{Kovalev2005,Taniguchi2014}. 

\section{Conclusion}\label{sec:conclusion}

In conclusion, we analyzed how the stability of a nanomagnet levitated in a static magnetic field is affected by the most relevant sources of dissipation. We find that in the strong-field regime (atom phase) the system is unstable due to the Gilbert-damping-induced magnetization switching, which occurs on a much faster timescale than the center-of-mass oscillations, thereby preventing the observation of levitation. On the other hand, the system is metastable in the weak-field regime and for small particle dimensions (Einstein--de Haas phase). In this regime, the confinement of the nanomagnet in a magnetic trap is limited only by the gas-induced dissipation. Our results suggest that the timescale of stable levitation can reach and even exceed several hundreds of periods of center-of-mass oscillations in high vacuum. These findings indicate the possibility of observing the phenomenon of quantum spin stabilized magnetic levitation, which we hope will encourage further experimental research. 

The analysis presented in this article is relevant for the community of levitated magnetic systems. Specifically, we give precise conditions for the observation of the phenomenon of quantum spin stabilized levitation under experimentally feasible conditions. Levitating a magnet in a time-independent gradient trap represents a new direction in the currently growing field of magnetic levitation of micro- and nanoparticles, which is interesting for two reasons. First, the experimental observation of stable magnetic levitation of a non-rotating nanomagnet would represent a direct observation of the quantum nature of magnetization. Second, the observation of such phenomenon would be a step towards controlling and using the rich physics of magnetically levitated nanomagnets, with applications in magnetometry and in tests of fundamental forces~\cite{Kimball2016,Millen2020,GonzalezBallestero2021,Stickler2021}.

\begin{acknowledgments}
We thank G. E. W. Bauer, J. J. García-Ripoll, O. Romero-Isart, and B.~A.~Stickler for helpful discussions. 
We are grateful to O.~Romero-Isart, B.~A.~Stickler and S.~Viola Kusminskiy for comments on an early version of the manuscript.
C.C.R.\ acknowledges funding from ERC Advanced Grant QENOCOBA under the EU Horizon 2020 program (Grant Agreement No. 742102). V.W. acknowledges funding from the Max Planck Society and from the Deutsche Forschungsgemeinschaft (DFG, German Research Foundation) through Project-ID 429529648-TRR 306 QuCoLiMa ("Quantum Cooperativity of Light and Matter"). A.E.R.L. thanks the AMS for the financial support. 
\end{acknowledgments}

\appendix

\section{Rotation to the body frame}\label{app:transformation}

In this appendix we define the transformation matrix between the body-fixed and the laboratory reference frames according to the ZYZ Euler angle convention, with the Euler angles denoted as $\Ovec = (\alpha,\beta,\gamma)^T$. We define the transformation between the laboratory frame $O\evec_x\evec_y\evec_z$ and the body frame $O\evec_1\evec_2\evec_3$ as follows, 
\be\label{lab-body-frame-transformation}
\begin{split}
	\begin{pmatrix}\evec_1\\\evec_2\\\evec_3\end{pmatrix} = 
	R(\Ovec)
	\begin{pmatrix}\evec_x\\\evec_y\\\evec_z\end{pmatrix},
\end{split}
\ee
where
\be\label{transformation-R}
\begin{split}
	R(\Ovec) \equiv R_z(\alpha)R_y(\beta)R_z(\gamma) = 
	\begin{pmatrix}
		\cos \gamma & \sin \gamma & 0\\
		-\sin \gamma & \cos \gamma & 0\\
		0 & 0 & 1
	\end{pmatrix}\\
	\begin{pmatrix}
		\cos \beta & 0 & -\sin \beta \\
		0 & 1 & 0\\
		-\sin \beta & 0 &  \cos \beta \\
	\end{pmatrix}
	\begin{pmatrix}
		\cos \alpha & \sin \alpha & 0\\
		-\sin \alpha & \cos \alpha & 0\\
		0 & 0 & 1
	\end{pmatrix}.
\end{split}
\ee
Accordingly, the components $v_j$ ($j = 1, 2, 3$) of a vector $\bm{v}$ in the body frame $O\evec_1\evec_2\evec_3$  and the components $v_\nu$ ($\nu = x, y, z$) of the same vector in the laboratory frame $O\evec_x\evec_y\evec_z$ are related as
\be\label{vector-transf}
\begin{split}
	\begin{pmatrix}v_1\\v_2\\v_3\end{pmatrix} = 
	R^T(\Ovec)
	\begin{pmatrix}v_x\\v_y\\v_z\end{pmatrix}.
\end{split}
\ee
The angular velocity of a rotating particle $\ovec$ can be written in terms of the Euler angles as $\ovec = \dot\alpha \evec_z + \dot\beta \evec_y' +\dot\gamma \evec_3$, where $(\evec_x',\evec_y',\evec_z')^T = R_z(\alpha)(\evec_x,\evec_y,\evec_z)^T$ denotes the frame $O\evec_x'\evec_y'\evec_z'$ obtained after the first rotation of the laboratory frame $O\evec_x\evec_y\evec_z$ in the ZYZ convention. By using \eqref{lab-body-frame-transformation} and \eqref{transformation-R}, we can rewrite angular velocity in terms of the body frame coordinates,  
\be\label{angular-velocity-body}
\begin{split}
	\ovec = 
	\dot\alpha \spare{R(\Ovec)^{-1} \begin{pmatrix}\evec_1\\\evec_2\\\evec_3\end{pmatrix}}_3 
	+ \dot\beta  \spare{R(\gamma)^{-1}\begin{pmatrix}\evec_1\\\evec_2\\\evec_3\end{pmatrix}}_2 
	+ \dot\gamma \evec_3, 
\end{split}
\ee
which is compactly written as $(\w_1,\w_2,\w_3)^T = A(\Ovec) \dotOvec$, with
\be\label{matrix-A}
A(\Ovec) = 
\begin{pmatrix}
	-\cos \gamma \sin \beta & \sin \gamma & 0  \\
	\sin \beta \sin \gamma  & \cos \gamma & 0 \\
	\cos \beta & 0 & 1\\
\end{pmatrix}.
\ee

\section{Dynamics in the presence of thermal fluctuations}\label{app:noise}

In this appendix we consider the dynamics of a levitated nanomagnet in the presence of stochastic forces and torques induced by the surrounding gas. The dynamics of the system are described by the following set of stochastic differential equations (SDE),
\begin{align}
	\text{d} \rtvec &= \wi \ptvec \text{d} t,\label{seom-r} \\
	\text{d} \evec_3  &= \ovec \times \evec_3\vphantom{\frac{1}{2}} \text{d}t,\label{seom-e}\\
	\text{d} \ptvec &= \spare{\wl \nabla_{\rtvec} [\mvec \cdot \bvec(\rtvec)]- \Gcm \ptvec}\text{d}t + \sqrt{D_\text{cm}} \text{d}\bm{W}_p,\label{seom-p}\\
	\text{d}\lvec  &= \spare{\wl \mvec \times \bvec(\rtvec) -\dotmvec - \Grot\lvec}\text{d}t+ \sqrt{D_\text{rot}} \text{d}\bm{W}_l, \label{seom-l}\\
	\text{d} \mvec  &= \frac{\mvec}{1+\etaG^2}\!  \times\! [\oveceff - \etaG\mvec\times\pare{\ovec+\oveceff+ \etaG \ovec \times \mvec  }]\text{d}t,\label{seom-m}
\end{align}
where we model the thermal fluctuations as uncorrelated Gaussian noise represented by a six-dimensional vector of independent Wiener increments $(\text{d} \bm{W}_p,\text{d} \bm{W}_l)^T$.
The corresponding diffusion rate is described by the 
tensors $D_\text{cm}$ and $D_\text{rot}$ which, in agreement with the fluctuation-dissipation theorem, are related to the corresponding dissipation tensors $\Gcm$ and $\Grot$ as $D_\text{cm}\equiv2\Gcm\chi, D_\text{rot}\equiv2\Grot\chi$, where $\chi \equiv M k_B T \gamma_0^2 a^2/\mu^2$. 

\begin{figure}
    \includegraphics[width=\columnwidth]{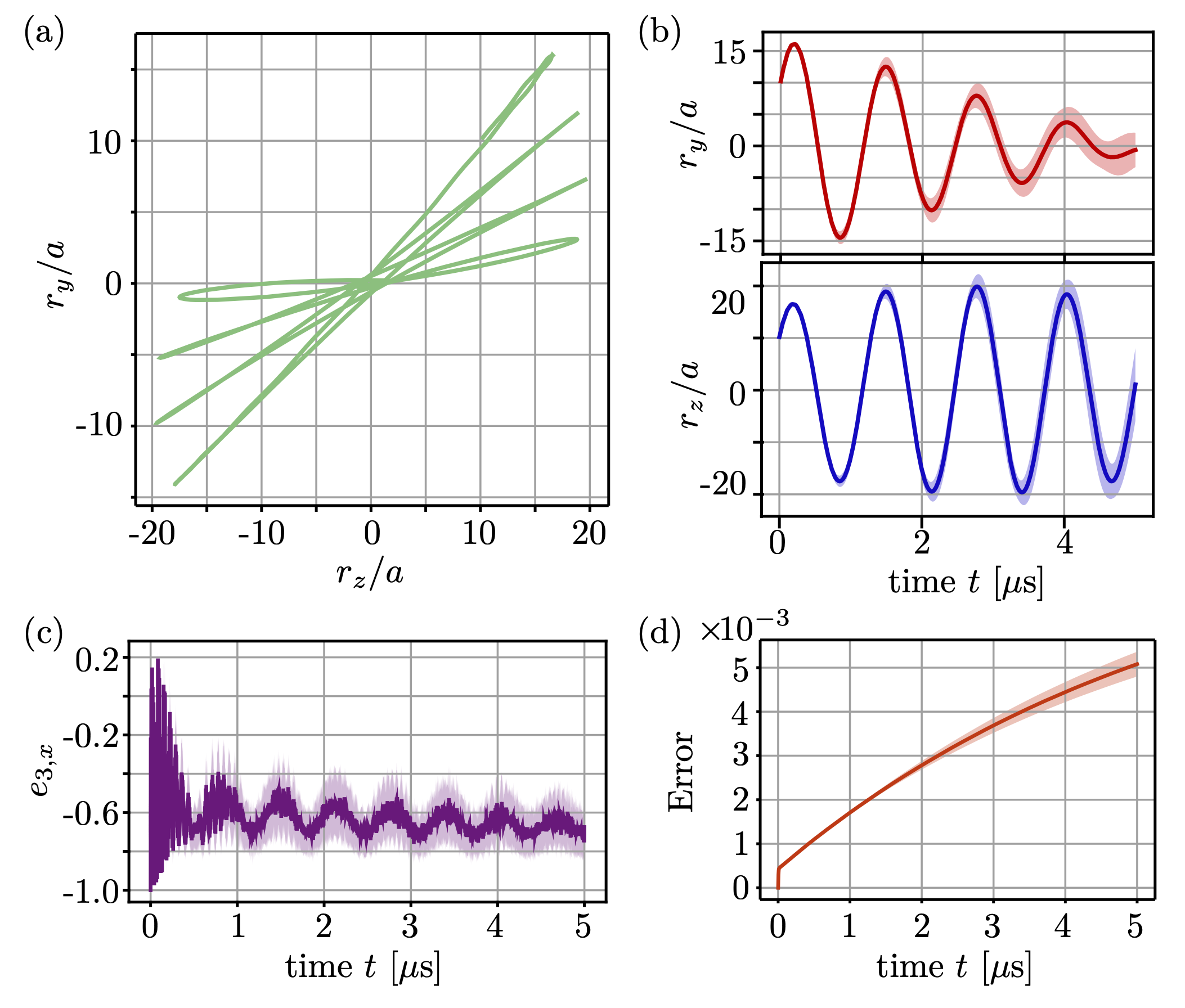}
    \caption{Stochastic dynamics of a nanomagnet for the same parameter regime as considered in~\figref{fig:EdH}.
    (a) Average motion of the system in the $y$-$z$ plane until time $t = 5$~$\mu$s.
    (b) Dynamics of center of mass along the $\evec_y$ (top) and $\evec_z$ (bottom) directions.
    (c) Dynamics of the anisotropy axis component $e_{3,x}$.
    (d) Numerical error as function of time.
    The simulations show the results of the average of $ 100$ different realizations of the system dynamics.
    In panels (b-d) the solid dark lines are the average trajectories, while the shaded area represents the standard deviation.}
    \label{fig:SDE_simulation}
\end{figure}

In the following we numerically integrate Eqs.~(\ref{seom-r}-\ref{seom-m}) using the stochastic Euler method implemented in the stochastic differential equations package in MATLAB. As the effect of thermal noise is more prominent for small particles at weak fields, we focus on the Einstein-de Haas regime considered in \secref{sec:EdH}. We show that even in this case the effect of thermal fluctuations leads to dynamics which are qualitatively very close to the results obtained in \secref{sec:EdH}. In \figref{fig:SDE_simulation} we present the results of the stochastic integrator by averaging the solution of 100 different trajectories calculated using the same parameters considered in \figref{fig:EdH}(a-c).
The resulting average dynamics agree qualitatively with the results obtained by integrating the corresponding set of deterministic equations Eqs.~(\ref{eom-r}-\ref{eom-m}) [cfr. \figref{fig:EdH}(a-c)].
The main effect of thermal excitations is to shift the center of oscillations of the particle's degrees of freedom around the value given by the thermal fluctuations. This is more evident for the dynamics of $\evec_3$ [cfr.~\figref{fig:SDE_simulation}(c) and \figref{fig:EdH}(c)].
We thus conclude that the deterministic equations Eqs.~(\ref{eom-r}-\ref{eom-m}) considered in the main text correctly capture the metastable behavior of the system. We emphasize that the results presented in this section include only the noise due to the surrounding gas. Should one be interested in simulating the effect of the fluctuations of the magnetic moment, the Euler method used here is not appropriate, and the Heun method should be used instead~\cite{Aron2014}. 

Let us conclude with a technical note on the numerical simulations. In the presence of dissipation and thermal fluctuations the only conserved quantity of the system is the magnitude of the magnetic moment ($|\mvec|=1$). We thus use the deviation $1-\vert\mvec\vert^2$ as a measure of the numerical error in both the stochastic and deterministic simulations presented in this article. For the deterministic simulations the error stays much smaller than any other physical degree of freedom of the system during the whole simulation time. The simulation of the stochastic dynamics shows a larger numerical error [see~\figref{fig:SDE_simulation}(d)], which can be partially reduced by taking a smaller time-step size. We note that, for the value of magnetic anisotropy given in \tabref{tab:parameters}, the system of SDE is stiff. This, together with the requirement imposed on the time-step size by the numerical error, ultimately limits the maximum time we can simulate to a few microseconds. However, this is sufficient to validate the agreement between the SDE and the deterministic simulations presented in the article.

\section{Additional figures}\label{app:additional}

\begin{figure*}
\centering
\begin{minipage}[t]{.45\textwidth}
    \centering
    \includegraphics[width=0.99\linewidth]{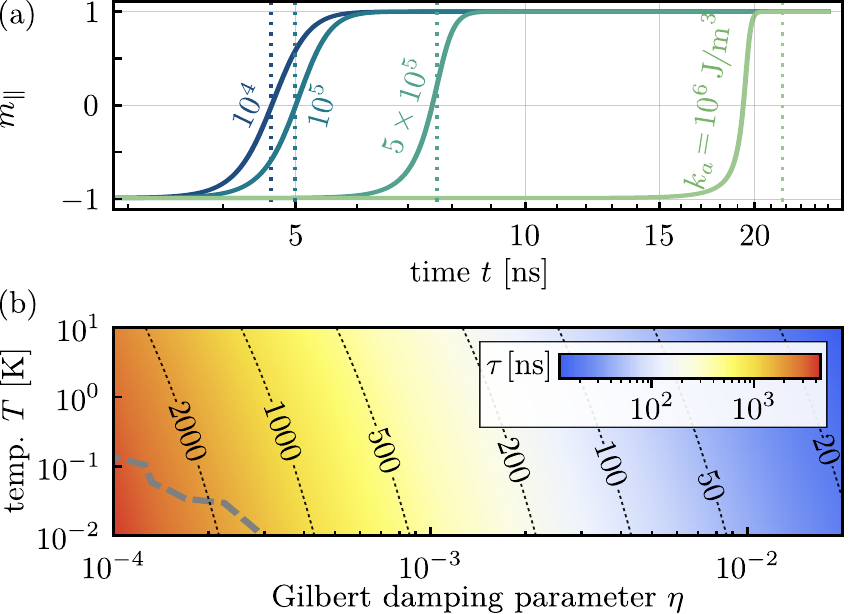}
    \caption{
        (a) Magnetization switching for different values of the anisotropy constant $k_a$ for nanomagnet dimensions $a=2b=20$ nm and the bias field $B_0=1100$ mT. For the initial conditions we consider trapping frequencies $\w_x = 2\pi \times \wxval$ kHz and $\w_y=\w_z = 2\pi \times \wtvala$ kHz. Unless otherwise stated, for the remaining parameters the numerical values are given in \tabref{tab:parameters}. Dotted vertical lines show \eqnref{tau}. 
        (b) Switching time given by \eqnref{tau} as a function of the Gilbert damping parameter $\eta$ and the temperature $T$ for nanomagnet dimensions $a=2b=20$ nm and the bias field $B_0=200$ mT, and the values of the remaining parameters same as in panel (a). In the region below the thick dashed line the deviation from the exact value is more than $5 \%$.} 
    \label{fig:atom-additional}
\end{minipage}
\hfill
\begin{minipage}[t]{.45\textwidth}
    \centering
    \includegraphics[width=0.99\linewidth]{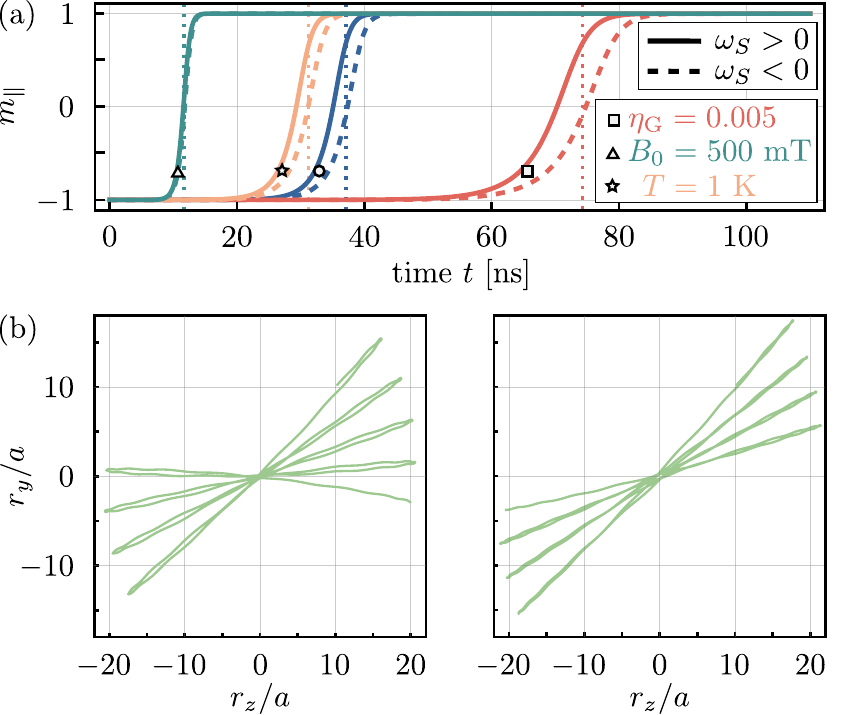}
    \caption{ 
        Dynamics of a nanomagnet initially rotating around the axis $\evec_x$ with frequency $\vert\w_S\vert/(2\pi) = 100$ MHz. 
        (a) Magnetization switching in the atom phase. Line denoted by circle corresponds to the same set of parameters as in \figref{fig:atom}(a). Each remaining line differs by a single parameter, as denoted by the legend. Dotted vertical lines show \eqnref{tau}. 
        (b) Motion in the $y$-$z$ plane in the Einstein--de Haas phase, using the same numerical values of the parameters as in \figref{fig:EdH}. Left panel: Clockwise rotation. Right panel: counterclockwise rotation. }
    \label{fig:rotation}
\end{minipage}
\end{figure*}

In this appendix we provide additional figures. 

\subsection{Dynamics in the atom phase}

In \figref{fig:atom-additional} we analyze magnetization dynamics in the atom phase as a function of different system parameters. In \figref{fig:atom-additional}(a) we show how magnetization switching changes as the anisotropy constant $k_a$ is varied. We consider the bias field $B_0=1100$ mT, which is larger than the value considered in the main text. This is done to ensure that $B_0 > B_\text{atom}$ for all anisotropy values. \figref{fig:atom-additional}(a) demonstrates that the switching time $\tau$, given by \eqnref{tau}, is an excellent approximation for the dynamics across a wide range of values for the anisotropy constant $k_a$. The larger discrepancy between \eqnref{tau} and the line showing the case with $k_a=10^6$ J/m$^3$ is explained by the proximity of this point to the unstable region (in this case given by the critical field $B_\text{atom} = 900$ mT), and better agreement is recovered at larger bias field values. 

In \figref{fig:atom-additional}(b) we analyze the validity of \eqnref{tau} for different values of the Gilbert damping parameter $\eta$ and the temperature $T$. The thick dashed line shows the region where \eqnref{tau} differs from the exact switching time by 5\%;  below this line the deviation becomes increasingly more significant. As evidenced by \figref{fig:atom-additional}(b), $\tau$ shows little dependence on $T$; its order of magnitude remains constant over a wide range of cryogenic temperatures. On the other hand, the dependence on $\eta$ is more pronounced. In fact, reducing the Gilbert parameter significantly delays the switching time, leading to levitation times as long as $\sim1$ $\mu$s. 

Additionally, we point out that $\tau$ depends on the field gradient $B'$ and curvature $B''$ through the initial condition $m_\parallel(t=0)$. In particular, magnetization switching can be delayed by decreasing $B'$, as this reduces the initial misalignment of the magnetization and the anisotropy axis (i. e. $\vert m_\parallel(t=0)\vert \to 1$).

\subsection{Dynamics in the presence of rotation}

In \figref{fig:rotation} we consider a more general equilibrium configuration, namely a nanomagnet initially rotating such that in the equilibrium point $\Lvec_e = -I_3\w_S \evec_x$, with $\w_S>0$ denoting the rotation in the clockwise direction. This equilibrium point is linearly stable in the absence of dissipation~\cite{Rusconi2017a,Rusconi2017b}, with additional stability of the system provided by the mechanical rotation, analogously to the classical magnetic top~\cite{Berry1996,Simon1997,Gov1999}.   

In \figref{fig:rotation}(a) we analyze how magnetization switching in the atom phase changes in the presence of rotation for different values of parameters. The rotation has a slight effect on the switching time $\tau$, shifting it forwards (backwards) in case of a clockwise (counterclockwise) rotation. This is generally expected in the presence of magneto-mechanical coupling~\cite{Kovalev2005,Taniguchi2014}. 

In \figref{fig:rotation}(b) we show the motion in the $y$-$z$ plane in the Einstein--de Haas phase for both directions of rotation. This can be compared with \figref{fig:EdH}(a). The rotation does not qualitatively affect the dynamics of the system. The difference in the two trajectories can be explained by a different total angular momentum in the two cases, as in the case of a clockwise (counterclockwise) rotation the mechanical and the internal angular momentum are parallel (anti-parallel), such that the total angular momentum is increased (decreased) compared to the non-rotating case. This asymmetry arises from the linear stability of a rotating nanomagnet, and it is not a consequence of dissipation. In fact, we confirm by numerical simulations that the escape time $t^\star$ as a function of the pressure $P$ shows no dependence on the mechanical rotation $\w_S$. Namely, even in the presence of mechanical rotation one recovers the same plot as shown in \figref{fig:EdH}(f).


\begin{thebibliography}{89}%
\makeatletter
\providecommand \@ifxundefined [1]{%
 \@ifx{#1\undefined}
}%
\providecommand \@ifnum [1]{%
 \ifnum #1\expandafter \@firstoftwo
 \else \expandafter \@secondoftwo
 \fi
}%
\providecommand \@ifx [1]{%
 \ifx #1\expandafter \@firstoftwo
 \else \expandafter \@secondoftwo
 \fi
}%
\providecommand \natexlab [1]{#1}%
\providecommand \enquote  [1]{``#1''}%
\providecommand \bibnamefont  [1]{#1}%
\providecommand \bibfnamefont [1]{#1}%
\providecommand \citenamefont [1]{#1}%
\providecommand \href@noop [0]{\@secondoftwo}%
\providecommand \href [0]{\begingroup \@sanitize@url \@href}%
\providecommand \@href[1]{\@@startlink{#1}\@@href}%
\providecommand \@@href[1]{\endgroup#1\@@endlink}%
\providecommand \@sanitize@url [0]{\catcode `\\12\catcode `\$12\catcode
  `\&12\catcode `\#12\catcode `\^12\catcode `\_12\catcode `\%12\relax}%
\providecommand \@@startlink[1]{}%
\providecommand \@@endlink[0]{}%
\providecommand \url  [0]{\begingroup\@sanitize@url \@url }%
\providecommand \@url [1]{\endgroup\@href {#1}{\urlprefix }}%
\providecommand \urlprefix  [0]{URL }%
\providecommand \Eprint [0]{\href }%
\providecommand \doibase [0]{https://doi.org/}%
\providecommand \selectlanguage [0]{\@gobble}%
\providecommand \bibinfo  [0]{\@secondoftwo}%
\providecommand \bibfield  [0]{\@secondoftwo}%
\providecommand \translation [1]{[#1]}%
\providecommand \BibitemOpen [0]{}%
\providecommand \bibitemStop [0]{}%
\providecommand \bibitemNoStop [0]{.\EOS\space}%
\providecommand \EOS [0]{\spacefactor3000\relax}%
\providecommand \BibitemShut  [1]{\csname bibitem#1\endcsname}%
\let\auto@bib@innerbib\@empty
\bibitem [{\citenamefont {{Einstein}}\ and\ \citenamefont {{de
  Haas}}(1915)}]{Einstein1915}%
  \BibitemOpen
  \bibfield  {author} {\bibinfo {author} {\bibfnamefont {A.}~\bibnamefont
  {{Einstein}}}\ and\ \bibinfo {author} {\bibfnamefont {W.~J.}\ \bibnamefont
  {{de Haas}}},\ }\bibfield  {title} {\bibinfo {title} {{Experimental proof of
  the existence of Amp{\`e}re's molecular currents}},\ }\href@noop {}
  {\bibfield  {journal} {\bibinfo  {journal} {Proc. K. Ned. Akad. Wet.}\
  }\textbf {\bibinfo {volume} {18}},\ \bibinfo {pages} {696} (\bibinfo {year}
  {1915})}\BibitemShut {NoStop}%
\bibitem [{\citenamefont {Richardson}(1908)}]{Richardson1908}%
  \BibitemOpen
  \bibfield  {author} {\bibinfo {author} {\bibfnamefont {O.~W.}\ \bibnamefont
  {Richardson}},\ }\bibfield  {title} {\bibinfo {title} {A mechanical effect
  accompanying magnetization},\ }\href
  {https://doi.org/10.1103/PhysRevSeriesI.26.248} {\bibfield  {journal}
  {\bibinfo  {journal} {Phys. Rev.}\ }\textbf {\bibinfo {volume} {26}},\
  \bibinfo {pages} {248} (\bibinfo {year} {1908})}\BibitemShut {NoStop}%
\bibitem [{\citenamefont {Barnett}(1915)}]{Barnett1915}%
  \BibitemOpen
  \bibfield  {author} {\bibinfo {author} {\bibfnamefont {S.~J.}\ \bibnamefont
  {Barnett}},\ }\bibfield  {title} {\bibinfo {title} {Magnetization by
  rotation},\ }\href {https://doi.org/10.1103/PhysRev.6.239} {\bibfield
  {journal} {\bibinfo  {journal} {Phys. Rev.}\ }\textbf {\bibinfo {volume}
  {6}},\ \bibinfo {pages} {239} (\bibinfo {year} {1915})}\BibitemShut {NoStop}%
\bibitem [{\citenamefont {Chudnovsky}(1994)}]{Chudnovsky1994}%
  \BibitemOpen
  \bibfield  {author} {\bibinfo {author} {\bibfnamefont {E.~M.}\ \bibnamefont
  {Chudnovsky}},\ }\bibfield  {title} {\bibinfo {title} {Conservation of
  angular momentum in the problem of tunneling of the magnetic moment},\ }\href
  {https://doi.org/10.1103/PhysRevLett.72.3433} {\bibfield  {journal} {\bibinfo
   {journal} {Phys. Rev. Lett.}\ }\textbf {\bibinfo {volume} {72}},\ \bibinfo
  {pages} {3433} (\bibinfo {year} {1994})}\BibitemShut {NoStop}%
\bibitem [{\citenamefont {Rusconi}\ and\ \citenamefont
  {Romero-Isart}(2016)}]{Rusconi2016}%
  \BibitemOpen
  \bibfield  {author} {\bibinfo {author} {\bibfnamefont {C.~C.}\ \bibnamefont
  {Rusconi}}\ and\ \bibinfo {author} {\bibfnamefont {O.}~\bibnamefont
  {Romero-Isart}},\ }\bibfield  {title} {\bibinfo {title} {Magnetic rigid rotor
  in the quantum regime: {Theoretical} toolbox},\ }\href
  {https://doi.org/10.1103/PhysRevB.93.054427} {\bibfield  {journal} {\bibinfo
  {journal} {Phys. Rev. B}\ }\textbf {\bibinfo {volume} {93}},\ \bibinfo
  {pages} {054427} (\bibinfo {year} {2016})}\BibitemShut {NoStop}%
\bibitem [{\citenamefont {Ganzhorn}\ \emph {et~al.}(2016)\citenamefont
  {Ganzhorn}, \citenamefont {Klyatskaya}, \citenamefont {Ruben},\ and\
  \citenamefont {Wernsdorfer}}]{Ganzhorn2016}%
  \BibitemOpen
  \bibfield  {author} {\bibinfo {author} {\bibfnamefont {M.}~\bibnamefont
  {Ganzhorn}}, \bibinfo {author} {\bibfnamefont {S.}~\bibnamefont
  {Klyatskaya}}, \bibinfo {author} {\bibfnamefont {M.}~\bibnamefont {Ruben}},\
  and\ \bibinfo {author} {\bibfnamefont {W.}~\bibnamefont {Wernsdorfer}},\
  }\bibfield  {title} {\bibinfo {title} {{Quantum Einstein-de Haas effect}},\
  }\href {https://doi.org/10.1038/ncomms11443} {\bibfield  {journal} {\bibinfo
  {journal} {Nat. Commun.}\ }\textbf {\bibinfo {volume} {7}},\ \bibinfo {pages}
  {11443} (\bibinfo {year} {2016})}\BibitemShut {NoStop}%
\bibitem [{\citenamefont {Viola~Kusminskiy}\ \emph {et~al.}(2016)\citenamefont
  {Viola~Kusminskiy}, \citenamefont {Tang},\ and\ \citenamefont
  {Marquardt}}]{Kusminskiy2016}%
  \BibitemOpen
  \bibfield  {author} {\bibinfo {author} {\bibfnamefont {S.}~\bibnamefont
  {Viola~Kusminskiy}}, \bibinfo {author} {\bibfnamefont {H.~X.}\ \bibnamefont
  {Tang}},\ and\ \bibinfo {author} {\bibfnamefont {F.}~\bibnamefont
  {Marquardt}},\ }\bibfield  {title} {\bibinfo {title} {Coupled spin-light
  dynamics in cavity optomagnonics},\ }\href
  {https://doi.org/10.1103/PhysRevA.94.033821} {\bibfield  {journal} {\bibinfo
  {journal} {Phys. Rev. A}\ }\textbf {\bibinfo {volume} {94}},\ \bibinfo
  {pages} {033821} (\bibinfo {year} {2016})}\BibitemShut {NoStop}%
\bibitem [{\citenamefont {Keshtgar}\ \emph {et~al.}(2017)\citenamefont
  {Keshtgar}, \citenamefont {Streib}, \citenamefont {Kamra}, \citenamefont
  {Blanter},\ and\ \citenamefont {Bauer}}]{Keshtgar2017}%
  \BibitemOpen
  \bibfield  {author} {\bibinfo {author} {\bibfnamefont {H.}~\bibnamefont
  {Keshtgar}}, \bibinfo {author} {\bibfnamefont {S.}~\bibnamefont {Streib}},
  \bibinfo {author} {\bibfnamefont {A.}~\bibnamefont {Kamra}}, \bibinfo
  {author} {\bibfnamefont {Y.~M.}\ \bibnamefont {Blanter}},\ and\ \bibinfo
  {author} {\bibfnamefont {G.~E.~W.}\ \bibnamefont {Bauer}},\ }\bibfield
  {title} {\bibinfo {title} {Magnetomechanical coupling and ferromagnetic
  resonance in magnetic nanoparticles},\ }\href
  {https://doi.org/10.1103/PhysRevB.95.134447} {\bibfield  {journal} {\bibinfo
  {journal} {Phys. Rev. B}\ }\textbf {\bibinfo {volume} {95}},\ \bibinfo
  {pages} {134447} (\bibinfo {year} {2017})}\BibitemShut {NoStop}%
\bibitem [{\citenamefont {Stickler}\ \emph {et~al.}(2021)\citenamefont
  {Stickler}, \citenamefont {Hornberger},\ and\ \citenamefont
  {Kim}}]{Stickler2021}%
  \BibitemOpen
  \bibfield  {author} {\bibinfo {author} {\bibfnamefont {B.~A.}\ \bibnamefont
  {Stickler}}, \bibinfo {author} {\bibfnamefont {K.}~\bibnamefont
  {Hornberger}},\ and\ \bibinfo {author} {\bibfnamefont {M.~S.}\ \bibnamefont
  {Kim}},\ }\bibfield  {title} {\bibinfo {title} {Quantum rotations of
  nanoparticles},\ }\href {https://doi.org/10.1038/s42254-021-00335-0}
  {\bibfield  {journal} {\bibinfo  {journal} {Nat. Rev. Phys.}\ }\textbf
  {\bibinfo {volume} {3}},\ \bibinfo {pages} {589} (\bibinfo {year}
  {2021})}\BibitemShut {NoStop}%
\bibitem [{\citenamefont {Perdriat}\ \emph {et~al.}(2021)\citenamefont
  {Perdriat}, \citenamefont {Pellet-Mary}, \citenamefont {Huillery},
  \citenamefont {Rondin},\ and\ \citenamefont {Hétet}}]{Perdriat2021}%
  \BibitemOpen
  \bibfield  {author} {\bibinfo {author} {\bibfnamefont {M.}~\bibnamefont
  {Perdriat}}, \bibinfo {author} {\bibfnamefont {C.}~\bibnamefont
  {Pellet-Mary}}, \bibinfo {author} {\bibfnamefont {P.}~\bibnamefont
  {Huillery}}, \bibinfo {author} {\bibfnamefont {L.}~\bibnamefont {Rondin}},\
  and\ \bibinfo {author} {\bibfnamefont {G.}~\bibnamefont {Hétet}},\
  }\bibfield  {title} {\bibinfo {title} {Spin-mechanics with nitrogen-vacancy
  centers and trapped particles},\ }\href {https://doi.org/10.3390/mi12060651}
  {\bibfield  {journal} {\bibinfo  {journal} {Micromachines}\ }\textbf
  {\bibinfo {volume} {12}},\ \bibinfo {pages} {651} (\bibinfo {year}
  {2021})}\BibitemShut {NoStop}%
\bibitem [{\citenamefont {Jackson~Kimball}\ \emph {et~al.}(2016)\citenamefont
  {Jackson~Kimball}, \citenamefont {Sushkov},\ and\ \citenamefont
  {Budker}}]{Kimball2016}%
  \BibitemOpen
  \bibfield  {author} {\bibinfo {author} {\bibfnamefont {D.~F.}\ \bibnamefont
  {Jackson~Kimball}}, \bibinfo {author} {\bibfnamefont {A.~O.}\ \bibnamefont
  {Sushkov}},\ and\ \bibinfo {author} {\bibfnamefont {D.}~\bibnamefont
  {Budker}},\ }\bibfield  {title} {\bibinfo {title} {Precessing ferromagnetic
  needle magnetometer},\ }\href
  {https://doi.org/10.1103/PhysRevLett.116.190801} {\bibfield  {journal}
  {\bibinfo  {journal} {Phys. Rev. Lett.}\ }\textbf {\bibinfo {volume} {116}},\
  \bibinfo {pages} {190801} (\bibinfo {year} {2016})}\BibitemShut {NoStop}%
\bibitem [{\citenamefont {Kumar}\ and\ \citenamefont
  {Bhattacharya}(2017)}]{Kumar2017}%
  \BibitemOpen
  \bibfield  {author} {\bibinfo {author} {\bibfnamefont {P.}~\bibnamefont
  {Kumar}}\ and\ \bibinfo {author} {\bibfnamefont {M.}~\bibnamefont
  {Bhattacharya}},\ }\bibfield  {title} {\bibinfo {title} {Magnetometry via
  spin-mechanical coupling in levitated optomechanics},\ }\href
  {https://doi.org/10.1364/OE.25.019568} {\bibfield  {journal} {\bibinfo
  {journal} {Opt. Express}\ }\textbf {\bibinfo {volume} {25}},\ \bibinfo
  {pages} {19568} (\bibinfo {year} {2017})}\BibitemShut {NoStop}%
\bibitem [{\citenamefont {Band}\ \emph {et~al.}(2018)\citenamefont {Band},
  \citenamefont {Avishai},\ and\ \citenamefont {Shnirman}}]{Band2018}%
  \BibitemOpen
  \bibfield  {author} {\bibinfo {author} {\bibfnamefont {Y.~B.}\ \bibnamefont
  {Band}}, \bibinfo {author} {\bibfnamefont {Y.}~\bibnamefont {Avishai}},\ and\
  \bibinfo {author} {\bibfnamefont {A.}~\bibnamefont {Shnirman}},\ }\bibfield
  {title} {\bibinfo {title} {Dynamics of a magnetic needle magnetometer:
  {Sensitivity} to {Landau-Lifshitz-Gilbert} damping},\ }\href
  {https://doi.org/10.1103/PhysRevLett.121.160801} {\bibfield  {journal}
  {\bibinfo  {journal} {Phys. Rev. Lett.}\ }\textbf {\bibinfo {volume} {121}},\
  \bibinfo {pages} {160801} (\bibinfo {year} {2018})}\BibitemShut {NoStop}%
\bibitem [{\citenamefont {Wang}\ \emph {et~al.}(2019)\citenamefont {Wang},
  \citenamefont {Lourette}, \citenamefont {O'Kelley}, \citenamefont {Kayci},
  \citenamefont {Band}, \citenamefont {Kimball}, \citenamefont {Sushkov},\ and\
  \citenamefont {Budker}}]{Wang2019}%
  \BibitemOpen
  \bibfield  {author} {\bibinfo {author} {\bibfnamefont {T.}~\bibnamefont
  {Wang}}, \bibinfo {author} {\bibfnamefont {S.}~\bibnamefont {Lourette}},
  \bibinfo {author} {\bibfnamefont {S.~R.}\ \bibnamefont {O'Kelley}}, \bibinfo
  {author} {\bibfnamefont {M.}~\bibnamefont {Kayci}}, \bibinfo {author}
  {\bibfnamefont {Y.}~\bibnamefont {Band}}, \bibinfo {author} {\bibfnamefont
  {D.~F.~J.}\ \bibnamefont {Kimball}}, \bibinfo {author} {\bibfnamefont
  {A.~O.}\ \bibnamefont {Sushkov}},\ and\ \bibinfo {author} {\bibfnamefont
  {D.}~\bibnamefont {Budker}},\ }\bibfield  {title} {\bibinfo {title} {Dynamics
  of a ferromagnetic particle levitated over a superconductor},\ }\href
  {https://doi.org/10.1103/PhysRevApplied.11.044041} {\bibfield  {journal}
  {\bibinfo  {journal} {Phys. Rev. Appl.}\ }\textbf {\bibinfo {volume} {11}},\
  \bibinfo {pages} {044041} (\bibinfo {year} {2019})}\BibitemShut {NoStop}%
\bibitem [{\citenamefont {Fadeev}\ \emph
  {et~al.}(2021{\natexlab{a}})\citenamefont {Fadeev}, \citenamefont
  {Timberlake}, \citenamefont {Wang}, \citenamefont {Vinante}, \citenamefont
  {Band}, \citenamefont {Budker}, \citenamefont {Sushkov}, \citenamefont
  {Ulbricht},\ and\ \citenamefont {Kimball}}]{Fadeev2021a}%
  \BibitemOpen
  \bibfield  {author} {\bibinfo {author} {\bibfnamefont {P.}~\bibnamefont
  {Fadeev}}, \bibinfo {author} {\bibfnamefont {C.}~\bibnamefont {Timberlake}},
  \bibinfo {author} {\bibfnamefont {T.}~\bibnamefont {Wang}}, \bibinfo {author}
  {\bibfnamefont {A.}~\bibnamefont {Vinante}}, \bibinfo {author} {\bibfnamefont
  {Y.~B.}\ \bibnamefont {Band}}, \bibinfo {author} {\bibfnamefont
  {D.}~\bibnamefont {Budker}}, \bibinfo {author} {\bibfnamefont {A.~O.}\
  \bibnamefont {Sushkov}}, \bibinfo {author} {\bibfnamefont {H.}~\bibnamefont
  {Ulbricht}},\ and\ \bibinfo {author} {\bibfnamefont {D.~F.~J.}\ \bibnamefont
  {Kimball}},\ }\bibfield  {title} {\bibinfo {title} {Ferromagnetic gyroscopes
  for tests of fundamental physics},\ }\href
  {https://doi.org/10.1088/2058-9565/abd892} {\bibfield  {journal} {\bibinfo
  {journal} {Quantum Sci. Technol.}\ }\textbf {\bibinfo {volume} {6}},\
  \bibinfo {pages} {024006} (\bibinfo {year} {2021}{\natexlab{a}})}\BibitemShut
  {NoStop}%
\bibitem [{\citenamefont {Fadeev}\ \emph
  {et~al.}(2021{\natexlab{b}})\citenamefont {Fadeev}, \citenamefont {Wang},
  \citenamefont {Band}, \citenamefont {Budker}, \citenamefont {Graham},
  \citenamefont {Sushkov},\ and\ \citenamefont {Kimball}}]{Fadeev2021b}%
  \BibitemOpen
  \bibfield  {author} {\bibinfo {author} {\bibfnamefont {P.}~\bibnamefont
  {Fadeev}}, \bibinfo {author} {\bibfnamefont {T.}~\bibnamefont {Wang}},
  \bibinfo {author} {\bibfnamefont {Y.~B.}\ \bibnamefont {Band}}, \bibinfo
  {author} {\bibfnamefont {D.}~\bibnamefont {Budker}}, \bibinfo {author}
  {\bibfnamefont {P.~W.}\ \bibnamefont {Graham}}, \bibinfo {author}
  {\bibfnamefont {A.~O.}\ \bibnamefont {Sushkov}},\ and\ \bibinfo {author}
  {\bibfnamefont {D.~F.~J.}\ \bibnamefont {Kimball}},\ }\bibfield  {title}
  {\bibinfo {title} {Gravity probe spin: {Prospects} for measuring
  general-relativistic precession of intrinsic spin using a ferromagnetic
  gyroscope},\ }\href {https://doi.org/10.1103/PhysRevD.103.044056} {\bibfield
  {journal} {\bibinfo  {journal} {Phys. Rev. D}\ }\textbf {\bibinfo {volume}
  {103}},\ \bibinfo {pages} {044056} (\bibinfo {year}
  {2021}{\natexlab{b}})}\BibitemShut {NoStop}%
\bibitem [{\citenamefont {Prat-Camps}\ \emph {et~al.}(2017)\citenamefont
  {Prat-Camps}, \citenamefont {Teo}, \citenamefont {Rusconi}, \citenamefont
  {Wieczorek},\ and\ \citenamefont {Romero-Isart}}]{PratCamps2017}%
  \BibitemOpen
  \bibfield  {author} {\bibinfo {author} {\bibfnamefont {J.}~\bibnamefont
  {Prat-Camps}}, \bibinfo {author} {\bibfnamefont {C.}~\bibnamefont {Teo}},
  \bibinfo {author} {\bibfnamefont {C.~C.}\ \bibnamefont {Rusconi}}, \bibinfo
  {author} {\bibfnamefont {W.}~\bibnamefont {Wieczorek}},\ and\ \bibinfo
  {author} {\bibfnamefont {O.}~\bibnamefont {Romero-Isart}},\ }\bibfield
  {title} {\bibinfo {title} {Ultrasensitive inertial and force sensors with
  diamagnetically levitated magnets},\ }\href
  {https://doi.org/10.1103/PhysRevApplied.8.034002} {\bibfield  {journal}
  {\bibinfo  {journal} {Phys. Rev. Appl.}\ }\textbf {\bibinfo {volume} {8}},\
  \bibinfo {pages} {034002} (\bibinfo {year} {2017})}\BibitemShut {NoStop}%
\bibitem [{\citenamefont {Vinante}\ \emph {et~al.}(2020)\citenamefont
  {Vinante}, \citenamefont {Falferi}, \citenamefont {Gasbarri}, \citenamefont
  {Setter}, \citenamefont {Timberlake},\ and\ \citenamefont
  {Ulbricht}}]{Vinante2020}%
  \BibitemOpen
  \bibfield  {author} {\bibinfo {author} {\bibfnamefont {A.}~\bibnamefont
  {Vinante}}, \bibinfo {author} {\bibfnamefont {P.}~\bibnamefont {Falferi}},
  \bibinfo {author} {\bibfnamefont {G.}~\bibnamefont {Gasbarri}}, \bibinfo
  {author} {\bibfnamefont {A.}~\bibnamefont {Setter}}, \bibinfo {author}
  {\bibfnamefont {C.}~\bibnamefont {Timberlake}},\ and\ \bibinfo {author}
  {\bibfnamefont {H.}~\bibnamefont {Ulbricht}},\ }\bibfield  {title} {\bibinfo
  {title} {Ultralow mechanical damping with {Meissner}-levitated ferromagnetic
  microparticles},\ }\href {https://doi.org/10.1103/PhysRevApplied.13.064027}
  {\bibfield  {journal} {\bibinfo  {journal} {Phys. Rev. Appl.}\ }\textbf
  {\bibinfo {volume} {13}},\ \bibinfo {pages} {064027} (\bibinfo {year}
  {2020})}\BibitemShut {NoStop}%
\bibitem [{\citenamefont {Huillery}\ \emph {et~al.}(2020)\citenamefont
  {Huillery}, \citenamefont {Delord}, \citenamefont {Nicolas}, \citenamefont
  {Van Den~Bossche}, \citenamefont {Perdriat},\ and\ \citenamefont
  {H\'etet}}]{Huillery2020}%
  \BibitemOpen
  \bibfield  {author} {\bibinfo {author} {\bibfnamefont {P.}~\bibnamefont
  {Huillery}}, \bibinfo {author} {\bibfnamefont {T.}~\bibnamefont {Delord}},
  \bibinfo {author} {\bibfnamefont {L.}~\bibnamefont {Nicolas}}, \bibinfo
  {author} {\bibfnamefont {M.}~\bibnamefont {Van Den~Bossche}}, \bibinfo
  {author} {\bibfnamefont {M.}~\bibnamefont {Perdriat}},\ and\ \bibinfo
  {author} {\bibfnamefont {G.}~\bibnamefont {H\'etet}},\ }\bibfield  {title}
  {\bibinfo {title} {Spin mechanics with levitating ferromagnetic particles},\
  }\href {https://doi.org/10.1103/PhysRevB.101.134415} {\bibfield  {journal}
  {\bibinfo  {journal} {Phys. Rev. B}\ }\textbf {\bibinfo {volume} {101}},\
  \bibinfo {pages} {134415} (\bibinfo {year} {2020})}\BibitemShut {NoStop}%
\bibitem [{\citenamefont {Gieseler}\ \emph {et~al.}(2020)\citenamefont
  {Gieseler}, \citenamefont {Kabcenell}, \citenamefont {Rosenfeld},
  \citenamefont {Schaefer}, \citenamefont {Safira}, \citenamefont {Schuetz},
  \citenamefont {Gonzalez-Ballestero}, \citenamefont {Rusconi}, \citenamefont
  {Romero-Isart},\ and\ \citenamefont {Lukin}}]{Gieseler2020}%
  \BibitemOpen
  \bibfield  {author} {\bibinfo {author} {\bibfnamefont {J.}~\bibnamefont
  {Gieseler}}, \bibinfo {author} {\bibfnamefont {A.}~\bibnamefont {Kabcenell}},
  \bibinfo {author} {\bibfnamefont {E.}~\bibnamefont {Rosenfeld}}, \bibinfo
  {author} {\bibfnamefont {J.~D.}\ \bibnamefont {Schaefer}}, \bibinfo {author}
  {\bibfnamefont {A.}~\bibnamefont {Safira}}, \bibinfo {author} {\bibfnamefont
  {M.~J.~A.}\ \bibnamefont {Schuetz}}, \bibinfo {author} {\bibfnamefont
  {C.}~\bibnamefont {Gonzalez-Ballestero}}, \bibinfo {author} {\bibfnamefont
  {C.~C.}\ \bibnamefont {Rusconi}}, \bibinfo {author} {\bibfnamefont
  {O.}~\bibnamefont {Romero-Isart}},\ and\ \bibinfo {author} {\bibfnamefont
  {M.~D.}\ \bibnamefont {Lukin}},\ }\bibfield  {title} {\bibinfo {title}
  {Single-spin magnetomechanics with levitated micromagnets},\ }\href
  {https://doi.org/10.1103/PhysRevLett.124.163604} {\bibfield  {journal}
  {\bibinfo  {journal} {Phys. Rev. Lett.}\ }\textbf {\bibinfo {volume} {124}},\
  \bibinfo {pages} {163604} (\bibinfo {year} {2020})}\BibitemShut {NoStop}%
\bibitem [{\citenamefont {Delord}\ \emph {et~al.}(2020)\citenamefont {Delord},
  \citenamefont {Huillery}, \citenamefont {Nicolas},\ and\ \citenamefont
  {H{\'e}tet}}]{Delord2020}%
  \BibitemOpen
  \bibfield  {author} {\bibinfo {author} {\bibfnamefont {T.}~\bibnamefont
  {Delord}}, \bibinfo {author} {\bibfnamefont {P.}~\bibnamefont {Huillery}},
  \bibinfo {author} {\bibfnamefont {L.}~\bibnamefont {Nicolas}},\ and\ \bibinfo
  {author} {\bibfnamefont {G.}~\bibnamefont {H{\'e}tet}},\ }\bibfield  {title}
  {\bibinfo {title} {Spin-cooling of the motion of a trapped diamond},\ }\href
  {https://doi.org/https://doi.org/10.1038/s41586-020-2133-z} {\bibfield
  {journal} {\bibinfo  {journal} {Nature}\ }\textbf {\bibinfo {volume} {580}},\
  \bibinfo {pages} {56} (\bibinfo {year} {2020})}\BibitemShut {NoStop}%
\bibitem [{\citenamefont {Gonzalez-Ballestero}\ \emph
  {et~al.}(2020)\citenamefont {Gonzalez-Ballestero}, \citenamefont {Gieseler},\
  and\ \citenamefont {Romero-Isart}}]{Gonzalez-Ballestero2020}%
  \BibitemOpen
  \bibfield  {author} {\bibinfo {author} {\bibfnamefont {C.}~\bibnamefont
  {Gonzalez-Ballestero}}, \bibinfo {author} {\bibfnamefont {J.}~\bibnamefont
  {Gieseler}},\ and\ \bibinfo {author} {\bibfnamefont {O.}~\bibnamefont
  {Romero-Isart}},\ }\bibfield  {title} {\bibinfo {title} {Quantum
  acoustomechanics with a micromagnet},\ }\href
  {https://doi.org/10.1103/PhysRevLett.124.093602} {\bibfield  {journal}
  {\bibinfo  {journal} {Phys. Rev. Lett.}\ }\textbf {\bibinfo {volume} {124}},\
  \bibinfo {pages} {093602} (\bibinfo {year} {2020})}\BibitemShut {NoStop}%
\bibitem [{\citenamefont {Rusconi}\ \emph
  {et~al.}(2017{\natexlab{a}})\citenamefont {Rusconi}, \citenamefont
  {P\"ochhacker}, \citenamefont {Kustura}, \citenamefont {Cirac},\ and\
  \citenamefont {Romero-Isart}}]{Rusconi2017a}%
  \BibitemOpen
  \bibfield  {author} {\bibinfo {author} {\bibfnamefont {C.~C.}\ \bibnamefont
  {Rusconi}}, \bibinfo {author} {\bibfnamefont {V.}~\bibnamefont
  {P\"ochhacker}}, \bibinfo {author} {\bibfnamefont {K.}~\bibnamefont
  {Kustura}}, \bibinfo {author} {\bibfnamefont {J.~I.}\ \bibnamefont {Cirac}},\
  and\ \bibinfo {author} {\bibfnamefont {O.}~\bibnamefont {Romero-Isart}},\
  }\bibfield  {title} {\bibinfo {title} {Quantum spin stabilized magnetic
  levitation},\ }\href {https://doi.org/10.1103/PhysRevLett.119.167202}
  {\bibfield  {journal} {\bibinfo  {journal} {Phys. Rev. Lett.}\ }\textbf
  {\bibinfo {volume} {119}},\ \bibinfo {pages} {167202} (\bibinfo {year}
  {2017}{\natexlab{a}})}\BibitemShut {NoStop}%
\bibitem [{\citenamefont {Rusconi}\ \emph
  {et~al.}(2017{\natexlab{b}})\citenamefont {Rusconi}, \citenamefont
  {P\"ochhacker}, \citenamefont {Cirac},\ and\ \citenamefont
  {Romero-Isart}}]{Rusconi2017b}%
  \BibitemOpen
  \bibfield  {author} {\bibinfo {author} {\bibfnamefont {C.~C.}\ \bibnamefont
  {Rusconi}}, \bibinfo {author} {\bibfnamefont {V.}~\bibnamefont
  {P\"ochhacker}}, \bibinfo {author} {\bibfnamefont {J.~I.}\ \bibnamefont
  {Cirac}},\ and\ \bibinfo {author} {\bibfnamefont {O.}~\bibnamefont
  {Romero-Isart}},\ }\bibfield  {title} {\bibinfo {title} {Linear stability
  analysis of a levitated nanomagnet in a static magnetic field: {Quantum} spin
  stabilized magnetic levitation},\ }\href
  {https://doi.org/10.1103/PhysRevB.96.134419} {\bibfield  {journal} {\bibinfo
  {journal} {Phys. Rev. B}\ }\textbf {\bibinfo {volume} {96}},\ \bibinfo
  {pages} {134419} (\bibinfo {year} {2017}{\natexlab{b}})}\BibitemShut
  {NoStop}%
\bibitem [{\citenamefont {Berry}(1996)}]{Berry1996}%
  \BibitemOpen
  \bibfield  {author} {\bibinfo {author} {\bibfnamefont {M.~V.}\ \bibnamefont
  {Berry}},\ }\bibfield  {title} {\bibinfo {title} {{The LevitronTM: an
  adiabatic trap for spins}},\ }\href
  {https://doi.org/https://doi.org/10.1098/rspa.1996.0062} {\bibfield
  {journal} {\bibinfo  {journal} {Proc. R. Soc. Lond. A}\ }\textbf {\bibinfo
  {volume} {452}},\ \bibinfo {pages} {1207} (\bibinfo {year}
  {1996})}\BibitemShut {NoStop}%
\bibitem [{\citenamefont {Simon}\ \emph {et~al.}(1997)\citenamefont {Simon},
  \citenamefont {Heflinger},\ and\ \citenamefont {Ridgway}}]{Simon1997}%
  \BibitemOpen
  \bibfield  {author} {\bibinfo {author} {\bibfnamefont {M.~D.}\ \bibnamefont
  {Simon}}, \bibinfo {author} {\bibfnamefont {L.~O.}\ \bibnamefont
  {Heflinger}},\ and\ \bibinfo {author} {\bibfnamefont {S.~L.}\ \bibnamefont
  {Ridgway}},\ }\bibfield  {title} {\bibinfo {title} {{Spin stabilized magnetic
  levitation}},\ }\href {https://doi.org/10.1119/1.18488} {\bibfield  {journal}
  {\bibinfo  {journal} {Am. J. Phys.}\ }\textbf {\bibinfo {volume} {65}},\
  \bibinfo {pages} {286} (\bibinfo {year} {1997})}\BibitemShut {NoStop}%
\bibitem [{\citenamefont {Gov}\ \emph {et~al.}(1999)\citenamefont {Gov},
  \citenamefont {Shtrikman},\ and\ \citenamefont {Thomas}}]{Gov1999}%
  \BibitemOpen
  \bibfield  {author} {\bibinfo {author} {\bibfnamefont {S.}~\bibnamefont
  {Gov}}, \bibinfo {author} {\bibfnamefont {S.}~\bibnamefont {Shtrikman}},\
  and\ \bibinfo {author} {\bibfnamefont {H.}~\bibnamefont {Thomas}},\
  }\bibfield  {title} {\bibinfo {title} {On the dynamical stability of the
  hovering magnetic top},\ }\href
  {https://doi.org/https://doi.org/10.1016/S0167-2789(98)00282-6} {\bibfield
  {journal} {\bibinfo  {journal} {Physica D}\ }\textbf {\bibinfo {volume}
  {126}},\ \bibinfo {pages} {214} (\bibinfo {year} {1999})}\BibitemShut
  {NoStop}%
\bibitem [{\citenamefont {Merkin}(2012)}]{Merkin2012}%
  \BibitemOpen
  \bibfield  {author} {\bibinfo {author} {\bibfnamefont {D.~R.}\ \bibnamefont
  {Merkin}},\ }\href@noop {} {\emph {\bibinfo {title} {Introduction to the
  Theory of Stability}}},\ Vol.~\bibinfo {volume} {24}\ (\bibinfo  {publisher}
  {Springer Science \& Business Media},\ \bibinfo {year} {2012})\BibitemShut
  {NoStop}%
\bibitem [{\citenamefont {Gilbert}(2004)}]{Gilbert2004}%
  \BibitemOpen
  \bibfield  {author} {\bibinfo {author} {\bibfnamefont {T.~L.}\ \bibnamefont
  {Gilbert}},\ }\bibfield  {title} {\bibinfo {title} {A phenomenological theory
  of damping in ferromagnetic materials},\ }\href
  {https://doi.org/10.1109/TMAG.2004.836740} {\bibfield  {journal} {\bibinfo
  {journal} {IEEE Transactions on Magnetics}\ }\textbf {\bibinfo {volume}
  {40}},\ \bibinfo {pages} {3443} (\bibinfo {year} {2004})}\BibitemShut
  {NoStop}%
\bibitem [{\citenamefont {Bertotti}\ \emph {et~al.}(2009)\citenamefont
  {Bertotti}, \citenamefont {Mayergoyz},\ and\ \citenamefont
  {Serpico}}]{Bertotti2009}%
  \BibitemOpen
  \bibfield  {author} {\bibinfo {author} {\bibfnamefont {G.}~\bibnamefont
  {Bertotti}}, \bibinfo {author} {\bibfnamefont {I.}~\bibnamefont
  {Mayergoyz}},\ and\ \bibinfo {author} {\bibfnamefont {C.}~\bibnamefont
  {Serpico}},\ }\href@noop {} {\emph {\bibinfo {title} {Nonlinear magnetization
  dynamics in nanosystems}}}\ (\bibinfo  {publisher} {Elsevier},\ \bibinfo
  {year} {2009})\BibitemShut {NoStop}%
\bibitem [{\citenamefont {Xi}\ \emph {et~al.}(2006)\citenamefont {Xi},
  \citenamefont {Gao}, \citenamefont {Shi},\ and\ \citenamefont
  {Xue}}]{Xi2006}%
  \BibitemOpen
  \bibfield  {author} {\bibinfo {author} {\bibfnamefont {H.}~\bibnamefont
  {Xi}}, \bibinfo {author} {\bibfnamefont {K.-Z.}\ \bibnamefont {Gao}},
  \bibinfo {author} {\bibfnamefont {Y.}~\bibnamefont {Shi}},\ and\ \bibinfo
  {author} {\bibfnamefont {S.}~\bibnamefont {Xue}},\ }\bibfield  {title}
  {\bibinfo {title} {Precessional dynamics of single-domain magnetic
  nanoparticles driven by small ac magnetic fields},\ }\href
  {https://doi.org/10.1088/0022-3727/39/22/002} {\bibfield  {journal} {\bibinfo
   {journal} {J. Phys. D: Appl. Phys.}\ }\textbf {\bibinfo {volume} {39}},\
  \bibinfo {pages} {4746} (\bibinfo {year} {2006})}\BibitemShut {NoStop}%
\bibitem [{\citenamefont {Martinetz}\ \emph {et~al.}(2018)\citenamefont
  {Martinetz}, \citenamefont {Hornberger},\ and\ \citenamefont
  {Stickler}}]{Martinetz2018}%
  \BibitemOpen
  \bibfield  {author} {\bibinfo {author} {\bibfnamefont {L.}~\bibnamefont
  {Martinetz}}, \bibinfo {author} {\bibfnamefont {K.}~\bibnamefont
  {Hornberger}},\ and\ \bibinfo {author} {\bibfnamefont {B.~A.}\ \bibnamefont
  {Stickler}},\ }\bibfield  {title} {\bibinfo {title} {Gas-induced friction and
  diffusion of rigid rotors},\ }\href
  {https://doi.org/10.1103/PhysRevE.97.052112} {\bibfield  {journal} {\bibinfo
  {journal} {Physical Review E}\ }\textbf {\bibinfo {volume} {97}},\ \bibinfo
  {pages} {052112} (\bibinfo {year} {2018})}\BibitemShut {NoStop}%
\bibitem [{\citenamefont {Lyutyy}\ \emph {et~al.}(2019)\citenamefont {Lyutyy},
  \citenamefont {Denisov},\ and\ \citenamefont {H\"anggi}}]{Lyutyy2019}%
  \BibitemOpen
  \bibfield  {author} {\bibinfo {author} {\bibfnamefont {T.~V.}\ \bibnamefont
  {Lyutyy}}, \bibinfo {author} {\bibfnamefont {S.~I.}\ \bibnamefont
  {Denisov}},\ and\ \bibinfo {author} {\bibfnamefont {P.}~\bibnamefont
  {H\"anggi}},\ }\bibfield  {title} {\bibinfo {title} {Dissipation-induced
  rotation of suspended ferromagnetic nanoparticles},\ }\href
  {https://doi.org/10.1103/PhysRevB.100.134403} {\bibfield  {journal} {\bibinfo
   {journal} {Phys. Rev. B}\ }\textbf {\bibinfo {volume} {100}},\ \bibinfo
  {pages} {134403} (\bibinfo {year} {2019})}\BibitemShut {NoStop}%
\bibitem [{\citenamefont {Millen}\ \emph {et~al.}(2020)\citenamefont {Millen},
  \citenamefont {Monteiro}, \citenamefont {Pettit},\ and\ \citenamefont
  {Vamivakas}}]{Millen2020}%
  \BibitemOpen
  \bibfield  {author} {\bibinfo {author} {\bibfnamefont {J.}~\bibnamefont
  {Millen}}, \bibinfo {author} {\bibfnamefont {T.~S.}\ \bibnamefont
  {Monteiro}}, \bibinfo {author} {\bibfnamefont {R.}~\bibnamefont {Pettit}},\
  and\ \bibinfo {author} {\bibfnamefont {A.~N.}\ \bibnamefont {Vamivakas}},\
  }\bibfield  {title} {\bibinfo {title} {Optomechanics with levitated
  particles},\ }\href {https://doi.org/10.1088/1361-6633/ab6100} {\bibfield
  {journal} {\bibinfo  {journal} {Rep. Prog. Phys.}\ }\textbf {\bibinfo
  {volume} {83}},\ \bibinfo {pages} {026401} (\bibinfo {year}
  {2020})}\BibitemShut {NoStop}%
\bibitem [{\citenamefont {Gonzalez-Ballestero}\ \emph
  {et~al.}(2021)\citenamefont {Gonzalez-Ballestero}, \citenamefont
  {Aspelmeyer}, \citenamefont {Novotny}, \citenamefont {Quidant},\ and\
  \citenamefont {Romero-Isart}}]{GonzalezBallestero2021}%
  \BibitemOpen
  \bibfield  {author} {\bibinfo {author} {\bibfnamefont {C.}~\bibnamefont
  {Gonzalez-Ballestero}}, \bibinfo {author} {\bibfnamefont {M.}~\bibnamefont
  {Aspelmeyer}}, \bibinfo {author} {\bibfnamefont {L.}~\bibnamefont {Novotny}},
  \bibinfo {author} {\bibfnamefont {R.}~\bibnamefont {Quidant}},\ and\ \bibinfo
  {author} {\bibfnamefont {O.}~\bibnamefont {Romero-Isart}},\ }\bibfield
  {title} {\bibinfo {title} {Levitodynamics: {Levitation} and control of
  microscopic objects in vacuum},\ }\href
  {https://doi.org/10.1126/science.abg3027} {\bibfield  {journal} {\bibinfo
  {journal} {Science}\ }\textbf {\bibinfo {volume} {374}},\ \bibinfo {pages}
  {eabg3027} (\bibinfo {year} {2021})}\BibitemShut {NoStop}%
\bibitem [{\citenamefont {Kuhlicke}\ \emph {et~al.}(2014)\citenamefont
  {Kuhlicke}, \citenamefont {Schell}, \citenamefont {Zoll},\ and\ \citenamefont
  {Benson}}]{Kuhlicke2014}%
  \BibitemOpen
  \bibfield  {author} {\bibinfo {author} {\bibfnamefont {A.}~\bibnamefont
  {Kuhlicke}}, \bibinfo {author} {\bibfnamefont {A.~W.}\ \bibnamefont
  {Schell}}, \bibinfo {author} {\bibfnamefont {J.}~\bibnamefont {Zoll}},\ and\
  \bibinfo {author} {\bibfnamefont {O.}~\bibnamefont {Benson}},\ }\bibfield
  {title} {\bibinfo {title} {Nitrogen vacancy center fluorescence from a
  submicron diamond cluster levitated in a linear quadrupole ion trap},\ }\href
  {https://doi.org/10.1063/1.4893575} {\bibfield  {journal} {\bibinfo
  {journal} {Applied Physics Letters}\ }\textbf {\bibinfo {volume} {105}},\
  \bibinfo {pages} {073101} (\bibinfo {year} {2014})}\BibitemShut {NoStop}%
\bibitem [{\citenamefont {Delord}\ \emph {et~al.}(2018)\citenamefont {Delord},
  \citenamefont {Huillery}, \citenamefont {Schwab}, \citenamefont {Nicolas},
  \citenamefont {Lecordier},\ and\ \citenamefont {H\'etet}}]{Delord2018}%
  \BibitemOpen
  \bibfield  {author} {\bibinfo {author} {\bibfnamefont {T.}~\bibnamefont
  {Delord}}, \bibinfo {author} {\bibfnamefont {P.}~\bibnamefont {Huillery}},
  \bibinfo {author} {\bibfnamefont {L.}~\bibnamefont {Schwab}}, \bibinfo
  {author} {\bibfnamefont {L.}~\bibnamefont {Nicolas}}, \bibinfo {author}
  {\bibfnamefont {L.}~\bibnamefont {Lecordier}},\ and\ \bibinfo {author}
  {\bibfnamefont {G.}~\bibnamefont {H\'etet}},\ }\bibfield  {title} {\bibinfo
  {title} {{Ramsey} interferences and spin echoes from electron spins inside a
  levitating macroscopic particle},\ }\href
  {https://doi.org/10.1103/PhysRevLett.121.053602} {\bibfield  {journal}
  {\bibinfo  {journal} {Phys. Rev. Lett.}\ }\textbf {\bibinfo {volume} {121}},\
  \bibinfo {pages} {053602} (\bibinfo {year} {2018})}\BibitemShut {NoStop}%
\bibitem [{\citenamefont {Slezak}\ \emph {et~al.}(2018)\citenamefont {Slezak},
  \citenamefont {Lewandowski}, \citenamefont {Hsu},\ and\ \citenamefont
  {D'Urso}}]{Slezak2018}%
  \BibitemOpen
  \bibfield  {author} {\bibinfo {author} {\bibfnamefont {B.~R.}\ \bibnamefont
  {Slezak}}, \bibinfo {author} {\bibfnamefont {C.~W.}\ \bibnamefont
  {Lewandowski}}, \bibinfo {author} {\bibfnamefont {J.-F.}\ \bibnamefont
  {Hsu}},\ and\ \bibinfo {author} {\bibfnamefont {B.}~\bibnamefont {D'Urso}},\
  }\bibfield  {title} {\bibinfo {title} {Cooling the motion of a silica
  microsphere in a magneto-gravitational trap in ultra-high vacuum},\ }\href
  {https://doi.org/10.1088/1367-2630/aacac1} {\bibfield  {journal} {\bibinfo
  {journal} {New Journal of Physics}\ }\textbf {\bibinfo {volume} {20}},\
  \bibinfo {pages} {063028} (\bibinfo {year} {2018})}\BibitemShut {NoStop}%
\bibitem [{\citenamefont {Zheng}\ \emph {et~al.}(2020)\citenamefont {Zheng},
  \citenamefont {Leng}, \citenamefont {Kong}, \citenamefont {Li}, \citenamefont
  {Wang}, \citenamefont {Luo}, \citenamefont {Zhao}, \citenamefont {Duan},
  \citenamefont {Huang}, \citenamefont {Du}, \citenamefont {Carlesso},\ and\
  \citenamefont {Bassi}}]{Zheng2020}%
  \BibitemOpen
  \bibfield  {author} {\bibinfo {author} {\bibfnamefont {D.}~\bibnamefont
  {Zheng}}, \bibinfo {author} {\bibfnamefont {Y.}~\bibnamefont {Leng}},
  \bibinfo {author} {\bibfnamefont {X.}~\bibnamefont {Kong}}, \bibinfo {author}
  {\bibfnamefont {R.}~\bibnamefont {Li}}, \bibinfo {author} {\bibfnamefont
  {Z.}~\bibnamefont {Wang}}, \bibinfo {author} {\bibfnamefont {X.}~\bibnamefont
  {Luo}}, \bibinfo {author} {\bibfnamefont {J.}~\bibnamefont {Zhao}}, \bibinfo
  {author} {\bibfnamefont {C.-K.}\ \bibnamefont {Duan}}, \bibinfo {author}
  {\bibfnamefont {P.}~\bibnamefont {Huang}}, \bibinfo {author} {\bibfnamefont
  {J.}~\bibnamefont {Du}}, \bibinfo {author} {\bibfnamefont {M.}~\bibnamefont
  {Carlesso}},\ and\ \bibinfo {author} {\bibfnamefont {A.}~\bibnamefont
  {Bassi}},\ }\bibfield  {title} {\bibinfo {title} {Room temperature test of
  the continuous spontaneous localization model using a levitated
  micro-oscillator},\ }\href {https://doi.org/10.1103/PhysRevResearch.2.013057}
  {\bibfield  {journal} {\bibinfo  {journal} {Phys. Rev. Research}\ }\textbf
  {\bibinfo {volume} {2}},\ \bibinfo {pages} {013057} (\bibinfo {year}
  {2020})}\BibitemShut {NoStop}%
\bibitem [{\citenamefont {Leng}\ \emph {et~al.}(2021)\citenamefont {Leng},
  \citenamefont {Li}, \citenamefont {Kong}, \citenamefont {Xie}, \citenamefont
  {Zheng}, \citenamefont {Yin}, \citenamefont {Xiong}, \citenamefont {Wu},
  \citenamefont {Duan}, \citenamefont {Du}, \citenamefont {Yin}, \citenamefont
  {Huang},\ and\ \citenamefont {Du}}]{Leng2021}%
  \BibitemOpen
  \bibfield  {author} {\bibinfo {author} {\bibfnamefont {Y.}~\bibnamefont
  {Leng}}, \bibinfo {author} {\bibfnamefont {R.}~\bibnamefont {Li}}, \bibinfo
  {author} {\bibfnamefont {X.}~\bibnamefont {Kong}}, \bibinfo {author}
  {\bibfnamefont {H.}~\bibnamefont {Xie}}, \bibinfo {author} {\bibfnamefont
  {D.}~\bibnamefont {Zheng}}, \bibinfo {author} {\bibfnamefont
  {P.}~\bibnamefont {Yin}}, \bibinfo {author} {\bibfnamefont {F.}~\bibnamefont
  {Xiong}}, \bibinfo {author} {\bibfnamefont {T.}~\bibnamefont {Wu}}, \bibinfo
  {author} {\bibfnamefont {C.-K.}\ \bibnamefont {Duan}}, \bibinfo {author}
  {\bibfnamefont {Y.}~\bibnamefont {Du}}, \bibinfo {author} {\bibfnamefont
  {Z.-q.}\ \bibnamefont {Yin}}, \bibinfo {author} {\bibfnamefont
  {P.}~\bibnamefont {Huang}},\ and\ \bibinfo {author} {\bibfnamefont
  {J.}~\bibnamefont {Du}},\ }\bibfield  {title} {\bibinfo {title} {Mechanical
  dissipation below $1\phantom{\rule{0.2em}{0ex}}\ensuremath{\mu}\mathrm{Hz}$
  with a cryogenic diamagnetic levitated micro-oscillator},\ }\href
  {https://doi.org/10.1103/PhysRevApplied.15.024061} {\bibfield  {journal}
  {\bibinfo  {journal} {Phys. Rev. Applied}\ }\textbf {\bibinfo {volume}
  {15}},\ \bibinfo {pages} {024061} (\bibinfo {year} {2021})}\BibitemShut
  {NoStop}%
\bibitem [{\citenamefont {Timberlake}\ \emph {et~al.}(2019)\citenamefont
  {Timberlake}, \citenamefont {Gasbarri}, \citenamefont {Vinante},
  \citenamefont {Setter},\ and\ \citenamefont {Ulbricht}}]{Timberlake2019}%
  \BibitemOpen
  \bibfield  {author} {\bibinfo {author} {\bibfnamefont {C.}~\bibnamefont
  {Timberlake}}, \bibinfo {author} {\bibfnamefont {G.}~\bibnamefont
  {Gasbarri}}, \bibinfo {author} {\bibfnamefont {A.}~\bibnamefont {Vinante}},
  \bibinfo {author} {\bibfnamefont {A.}~\bibnamefont {Setter}},\ and\ \bibinfo
  {author} {\bibfnamefont {H.}~\bibnamefont {Ulbricht}},\ }\bibfield  {title}
  {\bibinfo {title} {Acceleration sensing with magnetically levitated
  oscillators above a superconductor},\ }\href
  {https://doi.org/10.1063/1.5129145} {\bibfield  {journal} {\bibinfo
  {journal} {Appl. Phys. Lett.}\ }\textbf {\bibinfo {volume} {115}},\ \bibinfo
  {pages} {224101} (\bibinfo {year} {2019})}\BibitemShut {NoStop}%
\bibitem [{\citenamefont {Chikazumi}\ and\ \citenamefont
  {Graham}(2009)}]{Chikazumi2009}%
  \BibitemOpen
  \bibfield  {author} {\bibinfo {author} {\bibfnamefont {S.}~\bibnamefont
  {Chikazumi}}\ and\ \bibinfo {author} {\bibfnamefont {C.~D.}\ \bibnamefont
  {Graham}},\ }\href@noop {} {\emph {\bibinfo {title} {Physics of
  Ferromagnetism}}},\ Vol.~\bibinfo {volume} {94}\ (\bibinfo  {publisher}
  {Oxford University Press on Demand},\ \bibinfo {year} {2009})\BibitemShut
  {NoStop}%
\bibitem [{\citenamefont {Gatteschi}\ \emph {et~al.}(2006)\citenamefont
  {Gatteschi}, \citenamefont {Sessoli},\ and\ \citenamefont
  {Villain}}]{Gatteschi2006}%
  \BibitemOpen
  \bibfield  {author} {\bibinfo {author} {\bibfnamefont {D.}~\bibnamefont
  {Gatteschi}}, \bibinfo {author} {\bibfnamefont {R.}~\bibnamefont {Sessoli}},\
  and\ \bibinfo {author} {\bibfnamefont {J.}~\bibnamefont {Villain}},\
  }\href@noop {} {\emph {\bibinfo {title} {Molecular nanomagnets}}}\ (\bibinfo
  {publisher} {Oxford University Press},\ \bibinfo {year} {2006})\BibitemShut
  {NoStop}%
\bibitem [{\citenamefont {Newman}\ and\ \citenamefont
  {Yarbrough}(1968)}]{Newman1968}%
  \BibitemOpen
  \bibfield  {author} {\bibinfo {author} {\bibfnamefont {J.~J.}\ \bibnamefont
  {Newman}}\ and\ \bibinfo {author} {\bibfnamefont {R.~B.}\ \bibnamefont
  {Yarbrough}},\ }\bibfield  {title} {\bibinfo {title} {Motions of a magnetic
  particle in a viscous medium},\ }\href {https://doi.org/10.1063/1.1656014}
  {\bibfield  {journal} {\bibinfo  {journal} {J. Appl. Phys.}\ }\textbf
  {\bibinfo {volume} {39}},\ \bibinfo {pages} {5566} (\bibinfo {year}
  {1968})}\BibitemShut {NoStop}%
\bibitem [{\citenamefont {Tsebers}(1975)}]{Tsebers1975}%
  \BibitemOpen
  \bibfield  {author} {\bibinfo {author} {\bibfnamefont {A.}~\bibnamefont
  {Tsebers}},\ }\bibfield  {title} {\bibinfo {title} {Simultaneous rotational
  diffusion of the magnetic moment and the solid matrix of a single-domain
  ferromagnetic particle},\ }\href@noop {} {\bibfield  {journal} {\bibinfo
  {journal} {Magnetohydrodynamics}\ }\textbf {\bibinfo {volume} {11}},\
  \bibinfo {pages} {273} (\bibinfo {year} {1975})}\BibitemShut {NoStop}%
\bibitem [{\citenamefont {Scherer}\ and\ \citenamefont
  {Matuttis}(2000)}]{Scherer2000}%
  \BibitemOpen
  \bibfield  {author} {\bibinfo {author} {\bibfnamefont {C.}~\bibnamefont
  {Scherer}}\ and\ \bibinfo {author} {\bibfnamefont {H.-G.}\ \bibnamefont
  {Matuttis}},\ }\bibfield  {title} {\bibinfo {title} {Rotational dynamics of
  magnetic particles in suspensions},\ }\href
  {https://doi.org/10.1103/PhysRevE.63.011504} {\bibfield  {journal} {\bibinfo
  {journal} {Phys. Rev. E}\ }\textbf {\bibinfo {volume} {63}},\ \bibinfo
  {pages} {011504} (\bibinfo {year} {2000})}\BibitemShut {NoStop}%
\bibitem [{\citenamefont {Usadel}\ and\ \citenamefont
  {Usadel}(2015)}]{Usadel2015}%
  \BibitemOpen
  \bibfield  {author} {\bibinfo {author} {\bibfnamefont {K.~D.}\ \bibnamefont
  {Usadel}}\ and\ \bibinfo {author} {\bibfnamefont {C.}~\bibnamefont
  {Usadel}},\ }\bibfield  {title} {\bibinfo {title} {Dynamics of magnetic
  single domain particles embedded in a viscous liquid},\ }\href
  {https://doi.org/10.1063/1.4937919} {\bibfield  {journal} {\bibinfo
  {journal} {J. Appl. Phys.}\ }\textbf {\bibinfo {volume} {118}},\ \bibinfo
  {pages} {234303} (\bibinfo {year} {2015})}\BibitemShut {NoStop}%
\bibitem [{\citenamefont {Usov}\ and\ \citenamefont
  {Ya~Liubimov}(2015)}]{Usov2015}%
  \BibitemOpen
  \bibfield  {author} {\bibinfo {author} {\bibfnamefont {N.~A.}\ \bibnamefont
  {Usov}}\ and\ \bibinfo {author} {\bibfnamefont {B.}~\bibnamefont
  {Ya~Liubimov}},\ }\bibfield  {title} {\bibinfo {title} {Magnetic nanoparticle
  motion in external magnetic field},\ }\href
  {https://doi.org/https://doi.org/10.1016/j.jmmm.2015.03.035} {\bibfield
  {journal} {\bibinfo  {journal} {J. Magn. Magn. Mater.}\ }\textbf {\bibinfo
  {volume} {385}},\ \bibinfo {pages} {339} (\bibinfo {year}
  {2015})}\BibitemShut {NoStop}%
\bibitem [{\citenamefont {Lyutyy}\ \emph {et~al.}(2018)\citenamefont {Lyutyy},
  \citenamefont {Hryshko},\ and\ \citenamefont {Kovner}}]{Lyutyy2018}%
  \BibitemOpen
  \bibfield  {author} {\bibinfo {author} {\bibfnamefont {T.}~\bibnamefont
  {Lyutyy}}, \bibinfo {author} {\bibfnamefont {O.}~\bibnamefont {Hryshko}},\
  and\ \bibinfo {author} {\bibfnamefont {A.}~\bibnamefont {Kovner}},\
  }\bibfield  {title} {\bibinfo {title} {Power loss for a periodically driven
  ferromagnetic nanoparticle in a viscous fluid: The finite anisotropy
  aspects},\ }\href
  {https://doi.org/https://doi.org/10.1016/j.jmmm.2017.09.021} {\bibfield
  {journal} {\bibinfo  {journal} {Journal of Magnetism and Magnetic Materials}\
  }\textbf {\bibinfo {volume} {446}},\ \bibinfo {pages} {87} (\bibinfo {year}
  {2018})}\BibitemShut {NoStop}%
\bibitem [{\citenamefont {O'Keeffe}\ and\ \citenamefont
  {Chudnovsky}(2011)}]{OKeeffe2011}%
  \BibitemOpen
  \bibfield  {author} {\bibinfo {author} {\bibfnamefont {M.~F.}\ \bibnamefont
  {O'Keeffe}}\ and\ \bibinfo {author} {\bibfnamefont {E.~M.}\ \bibnamefont
  {Chudnovsky}},\ }\bibfield  {title} {\bibinfo {title} {Renormalization of the
  tunnel splitting in a rotating nanomagnet},\ }\href
  {https://doi.org/10.1103/PhysRevB.83.092402} {\bibfield  {journal} {\bibinfo
  {journal} {Phys. Rev. B}\ }\textbf {\bibinfo {volume} {83}},\ \bibinfo
  {pages} {092402} (\bibinfo {year} {2011})}\BibitemShut {NoStop}%
\bibitem [{\citenamefont {O'Keeffe}\ \emph {et~al.}(2012)\citenamefont
  {O'Keeffe}, \citenamefont {Chudnovsky},\ and\ \citenamefont
  {Garanin}}]{OKeeffe2012}%
  \BibitemOpen
  \bibfield  {author} {\bibinfo {author} {\bibfnamefont {M.~F.}\ \bibnamefont
  {O'Keeffe}}, \bibinfo {author} {\bibfnamefont {E.~M.}\ \bibnamefont
  {Chudnovsky}},\ and\ \bibinfo {author} {\bibfnamefont {D.~A.}\ \bibnamefont
  {Garanin}},\ }\bibfield  {title} {\bibinfo {title} {Quantum tunneling of the
  magnetic moment in a free nanoparticle},\ }\href
  {https://doi.org/https://doi.org/10.1016/j.jmmm.2012.04.032} {\bibfield
  {journal} {\bibinfo  {journal} {J. Magn. Magn. Mater.}\ }\textbf {\bibinfo
  {volume} {324}},\ \bibinfo {pages} {2871} (\bibinfo {year}
  {2012})}\BibitemShut {NoStop}%
\bibitem [{\citenamefont {Cercignani}(1988)}]{Cercignani1988}%
  \BibitemOpen
  \bibfield  {author} {\bibinfo {author} {\bibfnamefont {C.}~\bibnamefont
  {Cercignani}},\ }\bibinfo {title} {Small and large mean free paths},\ in\
  \href {https://doi.org/10.1007/978-1-4612-1039-9_5} {\emph {\bibinfo
  {booktitle} {The Boltzmann Equation and Its Applications}}}\ (\bibinfo
  {publisher} {Springer New York},\ \bibinfo {address} {New York, NY},\
  \bibinfo {year} {1988})\ pp.\ \bibinfo {pages} {232--285}\BibitemShut
  {NoStop}%
\bibitem [{\citenamefont {Miltat}\ \emph {et~al.}(2002)\citenamefont {Miltat},
  \citenamefont {Albuquerque},\ and\ \citenamefont {Thiaville}}]{Miltat2002}%
  \BibitemOpen
  \bibfield  {author} {\bibinfo {author} {\bibfnamefont {J.}~\bibnamefont
  {Miltat}}, \bibinfo {author} {\bibfnamefont {G.}~\bibnamefont
  {Albuquerque}},\ and\ \bibinfo {author} {\bibfnamefont {A.}~\bibnamefont
  {Thiaville}},\ }\bibinfo {title} {An introduction to micromagnetics in the
  dynamic regime},\ in\ \href {https://doi.org/10.1007/3-540-40907-6_1} {\emph
  {\bibinfo {booktitle} {Spin Dynamics in Confined Magnetic Structures I}}}\
  (\bibinfo  {publisher} {Springer Berlin Heidelberg},\ \bibinfo {year}
  {2002})\BibitemShut {NoStop}%
\bibitem [{\citenamefont {Brown}(1963)}]{Brown1963}%
  \BibitemOpen
  \bibfield  {author} {\bibinfo {author} {\bibfnamefont {W.~F.}\ \bibnamefont
  {Brown}},\ }\bibfield  {title} {\bibinfo {title} {Thermal fluctuations of a
  single-domain particle},\ }\href {https://doi.org/10.1103/PhysRev.130.1677}
  {\bibfield  {journal} {\bibinfo  {journal} {Phys. Rev.}\ }\textbf {\bibinfo
  {volume} {130}},\ \bibinfo {pages} {1677} (\bibinfo {year}
  {1963})}\BibitemShut {NoStop}%
\bibitem [{\citenamefont {Martinetz}\ \emph {et~al.}(2020)\citenamefont
  {Martinetz}, \citenamefont {Hornberger}, \citenamefont {Millen},
  \citenamefont {Kim},\ and\ \citenamefont {Stickler}}]{Martinetz2020}%
  \BibitemOpen
  \bibfield  {author} {\bibinfo {author} {\bibfnamefont {L.}~\bibnamefont
  {Martinetz}}, \bibinfo {author} {\bibfnamefont {K.}~\bibnamefont
  {Hornberger}}, \bibinfo {author} {\bibfnamefont {J.}~\bibnamefont {Millen}},
  \bibinfo {author} {\bibfnamefont {M.~S.}\ \bibnamefont {Kim}},\ and\ \bibinfo
  {author} {\bibfnamefont {B.~A.}\ \bibnamefont {Stickler}},\ }\bibfield
  {title} {\bibinfo {title} {Quantum electromechanics with levitated
  nanoparticles},\ }\href {https://doi.org/10.1038/s41534-020-00333-7}
  {\bibfield  {journal} {\bibinfo  {journal} {npj Quantum Inf.}\ }\textbf
  {\bibinfo {volume} {6}},\ \bibinfo {pages} {101} (\bibinfo {year}
  {2020})}\BibitemShut {NoStop}%
\bibitem [{\citenamefont {Sch\"afer}\ \emph {et~al.}(2021)\citenamefont
  {Sch\"afer}, \citenamefont {Rudolph}, \citenamefont {Hornberger},\ and\
  \citenamefont {Stickler}}]{Schaefer2021}%
  \BibitemOpen
  \bibfield  {author} {\bibinfo {author} {\bibfnamefont {J.}~\bibnamefont
  {Sch\"afer}}, \bibinfo {author} {\bibfnamefont {H.}~\bibnamefont {Rudolph}},
  \bibinfo {author} {\bibfnamefont {K.}~\bibnamefont {Hornberger}},\ and\
  \bibinfo {author} {\bibfnamefont {B.~A.}\ \bibnamefont {Stickler}},\
  }\bibfield  {title} {\bibinfo {title} {Cooling nanorotors by elliptic
  coherent scattering},\ }\href
  {https://doi.org/10.1103/PhysRevLett.126.163603} {\bibfield  {journal}
  {\bibinfo  {journal} {Phys. Rev. Lett.}\ }\textbf {\bibinfo {volume} {126}},\
  \bibinfo {pages} {163603} (\bibinfo {year} {2021})}\BibitemShut {NoStop}%
\bibitem [{\citenamefont {Taylor}(2005)}]{Taylor2005}%
  \BibitemOpen
  \bibfield  {author} {\bibinfo {author} {\bibfnamefont {J.}~\bibnamefont
  {Taylor}},\ }\href@noop {} {\emph {\bibinfo {title} {Classical mechanics}}}\
  (\bibinfo  {publisher} {University Science Books},\ \bibinfo {year}
  {2005})\BibitemShut {NoStop}%
\bibitem [{\citenamefont {Deli{\'c}}\ \emph {et~al.}(2020)\citenamefont
  {Deli{\'c}}, \citenamefont {Reisenbauer}, \citenamefont {Dare}, \citenamefont
  {Grass}, \citenamefont {Vuleti{\'c}}, \citenamefont {Kiesel},\ and\
  \citenamefont {Aspelmeyer}}]{Delic2020}%
  \BibitemOpen
  \bibfield  {author} {\bibinfo {author} {\bibfnamefont {U.}~\bibnamefont
  {Deli{\'c}}}, \bibinfo {author} {\bibfnamefont {M.}~\bibnamefont
  {Reisenbauer}}, \bibinfo {author} {\bibfnamefont {K.}~\bibnamefont {Dare}},
  \bibinfo {author} {\bibfnamefont {D.}~\bibnamefont {Grass}}, \bibinfo
  {author} {\bibfnamefont {V.}~\bibnamefont {Vuleti{\'c}}}, \bibinfo {author}
  {\bibfnamefont {N.}~\bibnamefont {Kiesel}},\ and\ \bibinfo {author}
  {\bibfnamefont {M.}~\bibnamefont {Aspelmeyer}},\ }\bibfield  {title}
  {\bibinfo {title} {Cooling of a levitated nanoparticle to the motional
  quantum ground state},\ }\href {https://doi.org/10.1126/science.aba3993}
  {\bibfield  {journal} {\bibinfo  {journal} {Science}\ }\textbf {\bibinfo
  {volume} {367}},\ \bibinfo {pages} {892} (\bibinfo {year}
  {2020})}\BibitemShut {NoStop}%
\bibitem [{\citenamefont {Magrini}\ \emph {et~al.}(2021)\citenamefont
  {Magrini}, \citenamefont {Rosenzweig}, \citenamefont {Bach}, \citenamefont
  {Deutschmann-Olek}, \citenamefont {Hofer}, \citenamefont {Hong},
  \citenamefont {Kiesel}, \citenamefont {Kugi},\ and\ \citenamefont
  {Aspelmeyer}}]{Magrini2021}%
  \BibitemOpen
  \bibfield  {author} {\bibinfo {author} {\bibfnamefont {L.}~\bibnamefont
  {Magrini}}, \bibinfo {author} {\bibfnamefont {P.}~\bibnamefont {Rosenzweig}},
  \bibinfo {author} {\bibfnamefont {C.}~\bibnamefont {Bach}}, \bibinfo {author}
  {\bibfnamefont {A.}~\bibnamefont {Deutschmann-Olek}}, \bibinfo {author}
  {\bibfnamefont {S.~G.}\ \bibnamefont {Hofer}}, \bibinfo {author}
  {\bibfnamefont {S.}~\bibnamefont {Hong}}, \bibinfo {author} {\bibfnamefont
  {N.}~\bibnamefont {Kiesel}}, \bibinfo {author} {\bibfnamefont
  {A.}~\bibnamefont {Kugi}},\ and\ \bibinfo {author} {\bibfnamefont
  {M.}~\bibnamefont {Aspelmeyer}},\ }\bibfield  {title} {\bibinfo {title}
  {Real-time optimal quantum control of mechanical motion at room
  temperature},\ }\href
  {https://doi.org/https://doi.org/10.1038/s41586-021-03602-3} {\bibfield
  {journal} {\bibinfo  {journal} {Nature}\ }\textbf {\bibinfo {volume} {595}},\
  \bibinfo {pages} {373} (\bibinfo {year} {2021})}\BibitemShut {NoStop}%
\bibitem [{\citenamefont {Tebbenjohanns}\ \emph {et~al.}(2021)\citenamefont
  {Tebbenjohanns}, \citenamefont {Mattana}, \citenamefont {Rossi},
  \citenamefont {Frimmer},\ and\ \citenamefont {Novotny}}]{Tebbenjohanns2021}%
  \BibitemOpen
  \bibfield  {author} {\bibinfo {author} {\bibfnamefont {F.}~\bibnamefont
  {Tebbenjohanns}}, \bibinfo {author} {\bibfnamefont {M.~L.}\ \bibnamefont
  {Mattana}}, \bibinfo {author} {\bibfnamefont {M.}~\bibnamefont {Rossi}},
  \bibinfo {author} {\bibfnamefont {M.}~\bibnamefont {Frimmer}},\ and\ \bibinfo
  {author} {\bibfnamefont {L.}~\bibnamefont {Novotny}},\ }\bibfield  {title}
  {\bibinfo {title} {Quantum control of a nanoparticle optically levitated in
  cryogenic free space},\ }\href
  {https://doi.org/https://doi.org/10.1038/s41586-021-03617-w} {\bibfield
  {journal} {\bibinfo  {journal} {Nature}\ }\textbf {\bibinfo {volume} {595}},\
  \bibinfo {pages} {378} (\bibinfo {year} {2021})}\BibitemShut {NoStop}%
\bibitem [{\citenamefont {Shliomis}(1974)}]{Shliomis1974}%
  \BibitemOpen
  \bibfield  {author} {\bibinfo {author} {\bibfnamefont {M.~I.}\ \bibnamefont
  {Shliomis}},\ }\bibfield  {title} {\bibinfo {title} {Magnetic fluids},\
  }\href {https://doi.org/10.1070/PU1974v017n02ABEH004332} {\bibfield
  {journal} {\bibinfo  {journal} {Sov. Phys.-Uspekhi}\ }\textbf {\bibinfo
  {volume} {17}},\ \bibinfo {pages} {153} (\bibinfo {year} {1974})}\BibitemShut
  {NoStop}%
\bibitem [{\citenamefont {Reichel}\ and\ \citenamefont
  {Vuletic}(2011)}]{Reichel2011}%
  \BibitemOpen
  \bibfield  {author} {\bibinfo {author} {\bibfnamefont {J.}~\bibnamefont
  {Reichel}}\ and\ \bibinfo {author} {\bibfnamefont {V.}~\bibnamefont
  {Vuletic}},\ }\href@noop {} {\emph {\bibinfo {title} {Atom chips}}}\
  (\bibinfo  {publisher} {John Wiley \& Sons},\ \bibinfo {year}
  {2011})\BibitemShut {NoStop}%
\bibitem [{\citenamefont {Millen}\ \emph {et~al.}(2015)\citenamefont {Millen},
  \citenamefont {Fonseca}, \citenamefont {Mavrogordatos}, \citenamefont
  {Monteiro},\ and\ \citenamefont {Barker}}]{Millen2015}%
  \BibitemOpen
  \bibfield  {author} {\bibinfo {author} {\bibfnamefont {J.}~\bibnamefont
  {Millen}}, \bibinfo {author} {\bibfnamefont {P.~Z.~G.}\ \bibnamefont
  {Fonseca}}, \bibinfo {author} {\bibfnamefont {T.}~\bibnamefont
  {Mavrogordatos}}, \bibinfo {author} {\bibfnamefont {T.~S.}\ \bibnamefont
  {Monteiro}},\ and\ \bibinfo {author} {\bibfnamefont {P.~F.}\ \bibnamefont
  {Barker}},\ }\bibfield  {title} {\bibinfo {title} {Cavity cooling a single
  charged levitated nanosphere},\ }\href
  {https://doi.org/10.1103/PhysRevLett.114.123602} {\bibfield  {journal}
  {\bibinfo  {journal} {Phys. Rev. Lett.}\ }\textbf {\bibinfo {volume} {114}},\
  \bibinfo {pages} {123602} (\bibinfo {year} {2015})}\BibitemShut {NoStop}%
\bibitem [{\citenamefont {Alda}\ \emph {et~al.}(2016)\citenamefont {Alda},
  \citenamefont {Berthelot}, \citenamefont {Rica},\ and\ \citenamefont
  {Quidant}}]{Alda2016}%
  \BibitemOpen
  \bibfield  {author} {\bibinfo {author} {\bibfnamefont {I.}~\bibnamefont
  {Alda}}, \bibinfo {author} {\bibfnamefont {J.}~\bibnamefont {Berthelot}},
  \bibinfo {author} {\bibfnamefont {R.~A.}\ \bibnamefont {Rica}},\ and\
  \bibinfo {author} {\bibfnamefont {R.}~\bibnamefont {Quidant}},\ }\bibfield
  {title} {\bibinfo {title} {Trapping and manipulation of individual
  nanoparticles in a planar {Paul} trap},\ }\href
  {https://doi.org/10.1063/1.4965859} {\bibfield  {journal} {\bibinfo
  {journal} {Appl. Phys. Lett.}\ }\textbf {\bibinfo {volume} {109}},\ \bibinfo
  {pages} {163105} (\bibinfo {year} {2016})}\BibitemShut {NoStop}%
\bibitem [{\citenamefont {Conangla}\ \emph {et~al.}(2018)\citenamefont
  {Conangla}, \citenamefont {Schell}, \citenamefont {Rica},\ and\ \citenamefont
  {Quidant}}]{Conangla2018}%
  \BibitemOpen
  \bibfield  {author} {\bibinfo {author} {\bibfnamefont {G.~P.}\ \bibnamefont
  {Conangla}}, \bibinfo {author} {\bibfnamefont {A.~W.}\ \bibnamefont
  {Schell}}, \bibinfo {author} {\bibfnamefont {R.~A.}\ \bibnamefont {Rica}},\
  and\ \bibinfo {author} {\bibfnamefont {R.}~\bibnamefont {Quidant}},\
  }\bibfield  {title} {\bibinfo {title} {Motion control and optical
  interrogation of a levitating single nitrogen vacancy in vacuum},\ }\href
  {https://doi.org/10.1021/acs.nanolett.8b01414} {\bibfield  {journal}
  {\bibinfo  {journal} {Nano Lett.}\ }\textbf {\bibinfo {volume} {18}},\
  \bibinfo {pages} {3956} (\bibinfo {year} {2018})}\BibitemShut {NoStop}%
\bibitem [{\citenamefont {Ostermayr}\ \emph {et~al.}(2018)\citenamefont
  {Ostermayr}, \citenamefont {Gebhard}, \citenamefont {Haffa}, \citenamefont
  {Kiefer}, \citenamefont {Kreuzer}, \citenamefont {Allinger}, \citenamefont
  {Bömer}, \citenamefont {Braenzel}, \citenamefont {Schnürer}, \citenamefont
  {Cermak}, \citenamefont {Schreiber},\ and\ \citenamefont
  {Hilz}}]{Ostermayr2018}%
  \BibitemOpen
  \bibfield  {author} {\bibinfo {author} {\bibfnamefont {T.~M.}\ \bibnamefont
  {Ostermayr}}, \bibinfo {author} {\bibfnamefont {J.}~\bibnamefont {Gebhard}},
  \bibinfo {author} {\bibfnamefont {D.}~\bibnamefont {Haffa}}, \bibinfo
  {author} {\bibfnamefont {D.}~\bibnamefont {Kiefer}}, \bibinfo {author}
  {\bibfnamefont {C.}~\bibnamefont {Kreuzer}}, \bibinfo {author} {\bibfnamefont
  {K.}~\bibnamefont {Allinger}}, \bibinfo {author} {\bibfnamefont
  {C.}~\bibnamefont {Bömer}}, \bibinfo {author} {\bibfnamefont
  {J.}~\bibnamefont {Braenzel}}, \bibinfo {author} {\bibfnamefont
  {M.}~\bibnamefont {Schnürer}}, \bibinfo {author} {\bibfnamefont
  {I.}~\bibnamefont {Cermak}}, \bibinfo {author} {\bibfnamefont
  {J.}~\bibnamefont {Schreiber}},\ and\ \bibinfo {author} {\bibfnamefont
  {P.}~\bibnamefont {Hilz}},\ }\bibfield  {title} {\bibinfo {title} {A
  transportable {Paul}-trap for levitation and accurate positioning of
  micron-scale particles in vacuum for laser-plasma experiments},\ }\href
  {https://doi.org/10.1063/1.4995955} {\bibfield  {journal} {\bibinfo
  {journal} {Rev. Sci. Instrum.}\ }\textbf {\bibinfo {volume} {89}},\ \bibinfo
  {pages} {013302} (\bibinfo {year} {2018})}\BibitemShut {NoStop}%
\bibitem [{\citenamefont {Partner}\ \emph {et~al.}(2018)\citenamefont
  {Partner}, \citenamefont {Zoll}, \citenamefont {Kuhlicke},\ and\
  \citenamefont {Benson}}]{Partner2018}%
  \BibitemOpen
  \bibfield  {author} {\bibinfo {author} {\bibfnamefont {H.~L.}\ \bibnamefont
  {Partner}}, \bibinfo {author} {\bibfnamefont {J.}~\bibnamefont {Zoll}},
  \bibinfo {author} {\bibfnamefont {A.}~\bibnamefont {Kuhlicke}},\ and\
  \bibinfo {author} {\bibfnamefont {O.}~\bibnamefont {Benson}},\ }\bibfield
  {title} {\bibinfo {title} {Printed-circuit-board linear {Paul} trap for
  manipulating single nano- and microparticles},\ }\href
  {https://doi.org/10.1063/1.5007924} {\bibfield  {journal} {\bibinfo
  {journal} {Rev. Sci. Instrum.}\ }\textbf {\bibinfo {volume} {89}},\ \bibinfo
  {pages} {083101} (\bibinfo {year} {2018})}\BibitemShut {NoStop}%
\bibitem [{\citenamefont {Bykov}\ \emph {et~al.}(2019)\citenamefont {Bykov},
  \citenamefont {Mestres}, \citenamefont {Dania}, \citenamefont {Schmöger},\
  and\ \citenamefont {Northup}}]{Bykov2019}%
  \BibitemOpen
  \bibfield  {author} {\bibinfo {author} {\bibfnamefont {D.~S.}\ \bibnamefont
  {Bykov}}, \bibinfo {author} {\bibfnamefont {P.}~\bibnamefont {Mestres}},
  \bibinfo {author} {\bibfnamefont {L.}~\bibnamefont {Dania}}, \bibinfo
  {author} {\bibfnamefont {L.}~\bibnamefont {Schmöger}},\ and\ \bibinfo
  {author} {\bibfnamefont {T.~E.}\ \bibnamefont {Northup}},\ }\bibfield
  {title} {\bibinfo {title} {Direct loading of nanoparticles under high vacuum
  into a {Paul} trap for levitodynamical experiments},\ }\href
  {https://doi.org/10.1063/1.5109645} {\bibfield  {journal} {\bibinfo
  {journal} {Appl. Phys. Lett.}\ }\textbf {\bibinfo {volume} {115}},\ \bibinfo
  {pages} {034101} (\bibinfo {year} {2019})}\BibitemShut {NoStop}%
\bibitem [{\citenamefont {Conangla}\ \emph {et~al.}(2020)\citenamefont
  {Conangla}, \citenamefont {Rica},\ and\ \citenamefont
  {Quidant}}]{Conangla2020}%
  \BibitemOpen
  \bibfield  {author} {\bibinfo {author} {\bibfnamefont {G.~P.}\ \bibnamefont
  {Conangla}}, \bibinfo {author} {\bibfnamefont {R.~A.}\ \bibnamefont {Rica}},\
  and\ \bibinfo {author} {\bibfnamefont {R.}~\bibnamefont {Quidant}},\
  }\bibfield  {title} {\bibinfo {title} {Extending vacuum trapping to absorbing
  objects with hybrid {Paul}-optical traps},\ }\href
  {https://doi.org/10.1021/acs.nanolett.0c02025} {\bibfield  {journal}
  {\bibinfo  {journal} {Nano Lett.}\ }\textbf {\bibinfo {volume} {20}},\
  \bibinfo {pages} {6018} (\bibinfo {year} {2020})}\BibitemShut {NoStop}%
\bibitem [{\citenamefont {Dania}\ \emph {et~al.}(2021)\citenamefont {Dania},
  \citenamefont {Bykov}, \citenamefont {Knoll}, \citenamefont {Mestres},\ and\
  \citenamefont {Northup}}]{Dania2021}%
  \BibitemOpen
  \bibfield  {author} {\bibinfo {author} {\bibfnamefont {L.}~\bibnamefont
  {Dania}}, \bibinfo {author} {\bibfnamefont {D.~S.}\ \bibnamefont {Bykov}},
  \bibinfo {author} {\bibfnamefont {M.}~\bibnamefont {Knoll}}, \bibinfo
  {author} {\bibfnamefont {P.}~\bibnamefont {Mestres}},\ and\ \bibinfo {author}
  {\bibfnamefont {T.~E.}\ \bibnamefont {Northup}},\ }\bibfield  {title}
  {\bibinfo {title} {Optical and electrical feedback cooling of a silica
  nanoparticle levitated in a {Paul} trap},\ }\href
  {https://doi.org/10.1103/PhysRevResearch.3.013018} {\bibfield  {journal}
  {\bibinfo  {journal} {Phys. Rev. Res.}\ }\textbf {\bibinfo {volume} {3}},\
  \bibinfo {pages} {013018} (\bibinfo {year} {2021})}\BibitemShut {NoStop}%
\bibitem [{\citenamefont {Martinetz}\ \emph {et~al.}(2021)\citenamefont
  {Martinetz}, \citenamefont {Hornberger},\ and\ \citenamefont
  {Stickler}}]{Martinetz2021}%
  \BibitemOpen
  \bibfield  {author} {\bibinfo {author} {\bibfnamefont {L.}~\bibnamefont
  {Martinetz}}, \bibinfo {author} {\bibfnamefont {K.}~\bibnamefont
  {Hornberger}},\ and\ \bibinfo {author} {\bibfnamefont {B.~A.}\ \bibnamefont
  {Stickler}},\ }\bibfield  {title} {\bibinfo {title} {Electric trapping and
  circuit cooling of charged nanorotors},\ }\href
  {https://doi.org/10.1088/1367-2630/ac1c82} {\bibfield  {journal} {\bibinfo
  {journal} {New Journal of Physics}\ }\textbf {\bibinfo {volume} {23}},\
  \bibinfo {pages} {093001} (\bibinfo {year} {2021})}\BibitemShut {NoStop}%
\bibitem [{\citenamefont {Sukumar}\ and\ \citenamefont
  {Brink}(1997)}]{Sukumar1997}%
  \BibitemOpen
  \bibfield  {author} {\bibinfo {author} {\bibfnamefont {C.~V.}\ \bibnamefont
  {Sukumar}}\ and\ \bibinfo {author} {\bibfnamefont {D.~M.}\ \bibnamefont
  {Brink}},\ }\bibfield  {title} {\bibinfo {title} {Spin-flip transitions in a
  magnetic trap},\ }\href {https://doi.org/10.1103/PhysRevA.56.2451} {\bibfield
   {journal} {\bibinfo  {journal} {Phys. Rev. A}\ }\textbf {\bibinfo {volume}
  {56}},\ \bibinfo {pages} {2451} (\bibinfo {year} {1997})}\BibitemShut
  {NoStop}%
\bibitem [{\citenamefont {Brink}\ and\ \citenamefont
  {Sukumar}(2006)}]{Brink2006}%
  \BibitemOpen
  \bibfield  {author} {\bibinfo {author} {\bibfnamefont {D.~M.}\ \bibnamefont
  {Brink}}\ and\ \bibinfo {author} {\bibfnamefont {C.~V.}\ \bibnamefont
  {Sukumar}},\ }\bibfield  {title} {\bibinfo {title} {Majorana spin-flip
  transitions in a magnetic trap},\ }\href
  {https://doi.org/10.1103/PhysRevA.74.035401} {\bibfield  {journal} {\bibinfo
  {journal} {Phys. Rev. A}\ }\textbf {\bibinfo {volume} {74}},\ \bibinfo
  {pages} {035401} (\bibinfo {year} {2006})}\BibitemShut {NoStop}%
\bibitem [{\citenamefont {Kovalev}\ \emph {et~al.}(2005)\citenamefont
  {Kovalev}, \citenamefont {Bauer},\ and\ \citenamefont
  {Brataas}}]{Kovalev2005}%
  \BibitemOpen
  \bibfield  {author} {\bibinfo {author} {\bibfnamefont {A.~A.}\ \bibnamefont
  {Kovalev}}, \bibinfo {author} {\bibfnamefont {G.~E.~W.}\ \bibnamefont
  {Bauer}},\ and\ \bibinfo {author} {\bibfnamefont {A.}~\bibnamefont
  {Brataas}},\ }\bibfield  {title} {\bibinfo {title} {Nanomechanical
  magnetization reversal},\ }\href
  {https://doi.org/10.1103/PhysRevLett.94.167201} {\bibfield  {journal}
  {\bibinfo  {journal} {Phys. Rev. Lett.}\ }\textbf {\bibinfo {volume} {94}},\
  \bibinfo {pages} {167201} (\bibinfo {year} {2005})}\BibitemShut {NoStop}%
\bibitem [{\citenamefont {Cullity}\ and\ \citenamefont
  {Graham}(2008)}]{Cullity2008}%
  \BibitemOpen
  \bibfield  {author} {\bibinfo {author} {\bibfnamefont {B.~D.}\ \bibnamefont
  {Cullity}}\ and\ \bibinfo {author} {\bibfnamefont {C.~D.}\ \bibnamefont
  {Graham}},\ }\href@noop {} {\emph {\bibinfo {title} {Introduction to magnetic
  materials}}}\ (\bibinfo  {publisher} {Wiley-IEEE Press},\ \bibinfo {year}
  {2008})\BibitemShut {NoStop}%
\bibitem [{\citenamefont {Walowski}\ \emph {et~al.}(2008)\citenamefont
  {Walowski}, \citenamefont {Djordjevic~Kaufmann}, \citenamefont {Lenk},
  \citenamefont {Hamann}, \citenamefont {McCord},\ and\ \citenamefont
  {Münzenberg}}]{Walowski2008}%
  \BibitemOpen
  \bibfield  {author} {\bibinfo {author} {\bibfnamefont {J.}~\bibnamefont
  {Walowski}}, \bibinfo {author} {\bibfnamefont {M.}~\bibnamefont
  {Djordjevic~Kaufmann}}, \bibinfo {author} {\bibfnamefont {B.}~\bibnamefont
  {Lenk}}, \bibinfo {author} {\bibfnamefont {C.}~\bibnamefont {Hamann}},
  \bibinfo {author} {\bibfnamefont {J.}~\bibnamefont {McCord}},\ and\ \bibinfo
  {author} {\bibfnamefont {M.}~\bibnamefont {Münzenberg}},\ }\bibfield
  {title} {\bibinfo {title} {Intrinsic and non-local {Gilbert} damping in
  polycrystalline nickel studied by
  {Ti}{\hspace{0.167em}}:{\hspace{0.167em}}sapphire laser fs spectroscopy},\
  }\href {https://doi.org/10.1088/0022-3727/41/16/164016} {\bibfield  {journal}
  {\bibinfo  {journal} {J. Phys. D: Appl. Phys.}\ }\textbf {\bibinfo {volume}
  {41}},\ \bibinfo {pages} {164016} (\bibinfo {year} {2008})}\BibitemShut
  {NoStop}%
\bibitem [{\citenamefont {Barati}\ \emph {et~al.}(2014)\citenamefont {Barati},
  \citenamefont {Cinal}, \citenamefont {Edwards},\ and\ \citenamefont
  {Umerski}}]{Barati2014}%
  \BibitemOpen
  \bibfield  {author} {\bibinfo {author} {\bibfnamefont {E.}~\bibnamefont
  {Barati}}, \bibinfo {author} {\bibfnamefont {M.}~\bibnamefont {Cinal}},
  \bibinfo {author} {\bibfnamefont {D.~M.}\ \bibnamefont {Edwards}},\ and\
  \bibinfo {author} {\bibfnamefont {A.}~\bibnamefont {Umerski}},\ }\bibfield
  {title} {\bibinfo {title} {{Gilbert} damping in magnetic layered systems},\
  }\href {https://doi.org/10.1103/PhysRevB.90.014420} {\bibfield  {journal}
  {\bibinfo  {journal} {Phys. Rev. B}\ }\textbf {\bibinfo {volume} {90}},\
  \bibinfo {pages} {014420} (\bibinfo {year} {2014})}\BibitemShut {NoStop}%
\bibitem [{\citenamefont {Papusoi}\ \emph {et~al.}(2018)\citenamefont
  {Papusoi}, \citenamefont {Le}, \citenamefont {Lo}, \citenamefont {Kaiser},
  \citenamefont {Desai},\ and\ \citenamefont {Acharya}}]{Papusoi2018}%
  \BibitemOpen
  \bibfield  {author} {\bibinfo {author} {\bibfnamefont {C.}~\bibnamefont
  {Papusoi}}, \bibinfo {author} {\bibfnamefont {T.}~\bibnamefont {Le}},
  \bibinfo {author} {\bibfnamefont {C.~C.~H.}\ \bibnamefont {Lo}}, \bibinfo
  {author} {\bibfnamefont {C.}~\bibnamefont {Kaiser}}, \bibinfo {author}
  {\bibfnamefont {M.}~\bibnamefont {Desai}},\ and\ \bibinfo {author}
  {\bibfnamefont {R.}~\bibnamefont {Acharya}},\ }\bibfield  {title} {\bibinfo
  {title} {{Measurements of Gilbert damping parameter $\alpha$ for {CoPt}-based
  and {CoFe}-based films for magnetic recording applications}},\ }\href
  {https://doi.org/10.1088/1361-6463/aacfcf} {\bibfield  {journal} {\bibinfo
  {journal} {J. Phys. D: Appl. Phys.}\ }\textbf {\bibinfo {volume} {51}},\
  \bibinfo {pages} {325002} (\bibinfo {year} {2018})}\BibitemShut {NoStop}%
\bibitem [{\citenamefont {Reichel}\ \emph {et~al.}(2001)\citenamefont
  {Reichel}, \citenamefont {H{\"a}nsel}, \citenamefont {Hommelhoff},\ and\
  \citenamefont {H{\"a}nsch}}]{Reichel2001}%
  \BibitemOpen
  \bibfield  {author} {\bibinfo {author} {\bibfnamefont {J.}~\bibnamefont
  {Reichel}}, \bibinfo {author} {\bibfnamefont {W.}~\bibnamefont {H{\"a}nsel}},
  \bibinfo {author} {\bibfnamefont {P.}~\bibnamefont {Hommelhoff}},\ and\
  \bibinfo {author} {\bibfnamefont {T.~W.}\ \bibnamefont {H{\"a}nsch}},\
  }\bibfield  {title} {\bibinfo {title} {Applications of integrated magnetic
  microtraps},\ }\href {https://doi.org/10.1007/s003400000460} {\bibfield
  {journal} {\bibinfo  {journal} {Appl. Phys. B}\ }\textbf {\bibinfo {volume}
  {72}},\ \bibinfo {pages} {81} (\bibinfo {year} {2001})}\BibitemShut {NoStop}%
\bibitem [{\citenamefont {Reichel}(2002)}]{Reichel2002}%
  \BibitemOpen
  \bibfield  {author} {\bibinfo {author} {\bibfnamefont {J.}~\bibnamefont
  {Reichel}},\ }\bibfield  {title} {\bibinfo {title} {Microchip traps and
  {Bose--Einstein} condensation},\ }\href
  {https://doi.org/10.1007/s003400200861} {\bibfield  {journal} {\bibinfo
  {journal} {Appl. Phys. B}\ }\textbf {\bibinfo {volume} {74}},\ \bibinfo
  {pages} {469} (\bibinfo {year} {2002})}\BibitemShut {NoStop}%
\bibitem [{\citenamefont {Barb}\ \emph {et~al.}(2005)\citenamefont {Barb},
  \citenamefont {Gerritsma}, \citenamefont {Xing}, \citenamefont {Goedkoop},\
  and\ \citenamefont {Spreeuw}}]{Barb2005}%
  \BibitemOpen
  \bibfield  {author} {\bibinfo {author} {\bibfnamefont {I.}~\bibnamefont
  {Barb}}, \bibinfo {author} {\bibfnamefont {R.}~\bibnamefont {Gerritsma}},
  \bibinfo {author} {\bibfnamefont {Y.~T.}\ \bibnamefont {Xing}}, \bibinfo
  {author} {\bibfnamefont {J.~B.}\ \bibnamefont {Goedkoop}},\ and\ \bibinfo
  {author} {\bibfnamefont {R.~J.~C.}\ \bibnamefont {Spreeuw}},\ }\bibfield
  {title} {\bibinfo {title} {Creating {Ioffe-Pritchard} micro-traps from
  permanent magnetic film with in-plane magnetization},\ }\href
  {https://doi.org/10.1140/epjd/e2005-00055-3} {\bibfield  {journal} {\bibinfo
  {journal} {Eur. Phys. J. D}\ }\textbf {\bibinfo {volume} {35}},\ \bibinfo
  {pages} {75} (\bibinfo {year} {2005})}\BibitemShut {NoStop}%
\bibitem [{\citenamefont {Fort\'agh}\ and\ \citenamefont
  {Zimmermann}(2007)}]{Fortagh2007}%
  \BibitemOpen
  \bibfield  {author} {\bibinfo {author} {\bibfnamefont {J.}~\bibnamefont
  {Fort\'agh}}\ and\ \bibinfo {author} {\bibfnamefont {C.}~\bibnamefont
  {Zimmermann}},\ }\bibfield  {title} {\bibinfo {title} {Magnetic microtraps
  for ultracold atoms},\ }\href {https://doi.org/10.1103/RevModPhys.79.235}
  {\bibfield  {journal} {\bibinfo  {journal} {Rev. Mod. Phys.}\ }\textbf
  {\bibinfo {volume} {79}},\ \bibinfo {pages} {235} (\bibinfo {year}
  {2007})}\BibitemShut {NoStop}%
\bibitem [{\citenamefont {Meyer}\ \emph {et~al.}(2019)\citenamefont {Meyer},
  \citenamefont {Sommer}, \citenamefont {Mestres}, \citenamefont {Gieseler},
  \citenamefont {Jain}, \citenamefont {Novotny},\ and\ \citenamefont
  {Quidant}}]{Meyer2019}%
  \BibitemOpen
  \bibfield  {author} {\bibinfo {author} {\bibfnamefont {N.}~\bibnamefont
  {Meyer}}, \bibinfo {author} {\bibfnamefont {A.~d. l.~R.}\ \bibnamefont
  {Sommer}}, \bibinfo {author} {\bibfnamefont {P.}~\bibnamefont {Mestres}},
  \bibinfo {author} {\bibfnamefont {J.}~\bibnamefont {Gieseler}}, \bibinfo
  {author} {\bibfnamefont {V.}~\bibnamefont {Jain}}, \bibinfo {author}
  {\bibfnamefont {L.}~\bibnamefont {Novotny}},\ and\ \bibinfo {author}
  {\bibfnamefont {R.}~\bibnamefont {Quidant}},\ }\bibfield  {title} {\bibinfo
  {title} {Resolved-sideband cooling of a levitated nanoparticle in the
  presence of laser phase noise},\ }\href
  {https://doi.org/10.1103/PhysRevLett.123.153601} {\bibfield  {journal}
  {\bibinfo  {journal} {Phys. Rev. Lett.}\ }\textbf {\bibinfo {volume} {123}},\
  \bibinfo {pages} {153601} (\bibinfo {year} {2019})}\BibitemShut {NoStop}%
\bibitem [{\citenamefont {Windey}\ \emph {et~al.}(2019)\citenamefont {Windey},
  \citenamefont {Gonzalez-Ballestero}, \citenamefont {Maurer}, \citenamefont
  {Novotny}, \citenamefont {Romero-Isart},\ and\ \citenamefont
  {Reimann}}]{Windey2019}%
  \BibitemOpen
  \bibfield  {author} {\bibinfo {author} {\bibfnamefont {D.}~\bibnamefont
  {Windey}}, \bibinfo {author} {\bibfnamefont {C.}~\bibnamefont
  {Gonzalez-Ballestero}}, \bibinfo {author} {\bibfnamefont {P.}~\bibnamefont
  {Maurer}}, \bibinfo {author} {\bibfnamefont {L.}~\bibnamefont {Novotny}},
  \bibinfo {author} {\bibfnamefont {O.}~\bibnamefont {Romero-Isart}},\ and\
  \bibinfo {author} {\bibfnamefont {R.}~\bibnamefont {Reimann}},\ }\bibfield
  {title} {\bibinfo {title} {Cavity-based {3D} cooling of a levitated
  nanoparticle via coherent scattering},\ }\href
  {https://doi.org/10.1103/PhysRevLett.122.123601} {\bibfield  {journal}
  {\bibinfo  {journal} {Phys. Rev. Lett.}\ }\textbf {\bibinfo {volume} {122}},\
  \bibinfo {pages} {123601} (\bibinfo {year} {2019})}\BibitemShut {NoStop}%
\bibitem [{\citenamefont {de~los R{\'\i}os~Sommer}\ \emph
  {et~al.}(2021)\citenamefont {de~los R{\'\i}os~Sommer}, \citenamefont
  {Meyer},\ and\ \citenamefont {Quidant}}]{Sommer2021}%
  \BibitemOpen
  \bibfield  {author} {\bibinfo {author} {\bibfnamefont {A.}~\bibnamefont
  {de~los R{\'\i}os~Sommer}}, \bibinfo {author} {\bibfnamefont
  {N.}~\bibnamefont {Meyer}},\ and\ \bibinfo {author} {\bibfnamefont
  {R.}~\bibnamefont {Quidant}},\ }\bibfield  {title} {\bibinfo {title} {Strong
  optomechanical coupling at room temperature by coherent scattering},\ }\href
  {https://doi.org/10.1038/s41467-020-20419-2} {\bibfield  {journal} {\bibinfo
  {journal} {Nat. Commun.}\ }\textbf {\bibinfo {volume} {12}},\ \bibinfo
  {pages} {276} (\bibinfo {year} {2021})}\BibitemShut {NoStop}%
\bibitem [{\citenamefont {van~der Laan}\ \emph {et~al.}(2020)\citenamefont
  {van~der Laan}, \citenamefont {Reimann}, \citenamefont {Militaru},
  \citenamefont {Tebbenjohanns}, \citenamefont {Windey}, \citenamefont
  {Frimmer},\ and\ \citenamefont {Novotny}}]{VanderLaan2020}%
  \BibitemOpen
  \bibfield  {author} {\bibinfo {author} {\bibfnamefont {F.}~\bibnamefont
  {van~der Laan}}, \bibinfo {author} {\bibfnamefont {R.}~\bibnamefont
  {Reimann}}, \bibinfo {author} {\bibfnamefont {A.}~\bibnamefont {Militaru}},
  \bibinfo {author} {\bibfnamefont {F.}~\bibnamefont {Tebbenjohanns}}, \bibinfo
  {author} {\bibfnamefont {D.}~\bibnamefont {Windey}}, \bibinfo {author}
  {\bibfnamefont {M.}~\bibnamefont {Frimmer}},\ and\ \bibinfo {author}
  {\bibfnamefont {L.}~\bibnamefont {Novotny}},\ }\bibfield  {title} {\bibinfo
  {title} {Optically levitated rotor at its thermal limit of frequency
  stability},\ }\href {https://doi.org/10.1103/PhysRevA.102.013505} {\bibfield
  {journal} {\bibinfo  {journal} {Phys. Rev. A}\ }\textbf {\bibinfo {volume}
  {102}},\ \bibinfo {pages} {013505} (\bibinfo {year} {2020})}\BibitemShut
  {NoStop}%
\bibitem [{\citenamefont {van~der Laan}\ \emph {et~al.}(2021)\citenamefont
  {van~der Laan}, \citenamefont {Tebbenjohanns}, \citenamefont {Reimann},
  \citenamefont {Vijayan}, \citenamefont {Novotny},\ and\ \citenamefont
  {Frimmer}}]{VanderLaan2021}%
  \BibitemOpen
  \bibfield  {author} {\bibinfo {author} {\bibfnamefont {F.}~\bibnamefont
  {van~der Laan}}, \bibinfo {author} {\bibfnamefont {F.}~\bibnamefont
  {Tebbenjohanns}}, \bibinfo {author} {\bibfnamefont {R.}~\bibnamefont
  {Reimann}}, \bibinfo {author} {\bibfnamefont {J.}~\bibnamefont {Vijayan}},
  \bibinfo {author} {\bibfnamefont {L.}~\bibnamefont {Novotny}},\ and\ \bibinfo
  {author} {\bibfnamefont {M.}~\bibnamefont {Frimmer}},\ }\bibfield  {title}
  {\bibinfo {title} {Sub-{Kelvin} feedback cooling and heating dynamics of an
  optically levitated librator},\ }\href
  {https://doi.org/10.1103/PhysRevLett.127.123605} {\bibfield  {journal}
  {\bibinfo  {journal} {Phys. Rev. Lett.}\ }\textbf {\bibinfo {volume} {127}},\
  \bibinfo {pages} {123605} (\bibinfo {year} {2021})}\BibitemShut {NoStop}%
\bibitem [{\citenamefont {Taniguchi}(2014)}]{Taniguchi2014}%
  \BibitemOpen
  \bibfield  {author} {\bibinfo {author} {\bibfnamefont {T.}~\bibnamefont
  {Taniguchi}},\ }\bibfield  {title} {\bibinfo {title} {Magnetization reversal
  condition for a nanomagnet within a rotating magnetic field},\ }\href
  {https://doi.org/10.1103/PhysRevB.90.024424} {\bibfield  {journal} {\bibinfo
  {journal} {Phys. Rev. B}\ }\textbf {\bibinfo {volume} {90}},\ \bibinfo
  {pages} {024424} (\bibinfo {year} {2014})}\BibitemShut {NoStop}%
\bibitem [{\citenamefont {Aron}\ \emph {et~al.}(2014)\citenamefont {Aron},
  \citenamefont {Barci}, \citenamefont {Cugliandolo}, \citenamefont {Arenas},\
  and\ \citenamefont {Lozano}}]{Aron2014}%
  \BibitemOpen
  \bibfield  {author} {\bibinfo {author} {\bibfnamefont {C.}~\bibnamefont
  {Aron}}, \bibinfo {author} {\bibfnamefont {D.~G.}\ \bibnamefont {Barci}},
  \bibinfo {author} {\bibfnamefont {L.~F.}\ \bibnamefont {Cugliandolo}},
  \bibinfo {author} {\bibfnamefont {Z.~G.}\ \bibnamefont {Arenas}},\ and\
  \bibinfo {author} {\bibfnamefont {G.~S.}\ \bibnamefont {Lozano}},\ }\bibfield
   {title} {\bibinfo {title} {{Magnetization dynamics: path-integral formalism
  for the stochastic Landau{\textendash}Lifshitz{\textendash}Gilbert
  equation}},\ }\href {https://doi.org/10.1088/1742-5468/2014/09/p09008}
  {\bibfield  {journal} {\bibinfo  {journal} {J. Stat. Mech.}\ ,\ \bibinfo
  {pages} {P09008}} (\bibinfo {year} {2014})}\BibitemShut {NoStop}%
\end{thebibliography}
\end{document}